\documentclass{article}

\usepackage[T1]{fontenc} % Modern font encoding
\usepackage{float}       % For creating charts, graphs and schemes
\usepackage{helvet}      % Helvetica font for sans serif
\usepackage{mathptmx}    % Times font ("Word-like")
\usepackage{setspace}    % For double-spacing

\usepackage{bm}
\usepackage{graphicx}
\usepackage{amssymb}
\usepackage{amsmath}
\usepackage{float}
\usepackage{xcolor}

\begin{document}

\title{\bf Using Pressure to Unravel the Structure-Dynamic-Disorder Relationship in Metal Halide Perovskites}

\maketitle

%\begin{document}

% Author: Please give full first and last names for authors and include * after the name of all corresponding authors

%\author{Author One}
%\author{Author Two}
%\author{Author Three*}
\centerline{Kai Xu$^{1}$, Luis P\'{e}rez-Fidalgo$^{1}$, Bethan L. Charles$^{2,3}$, Mark T.~Weller$^{4,5}$, M.~Isabel Alonso$^{1}$, and Alejandro R.~Go\~ni*$^{1,6}$}

\vspace{1cm}
% Dedication
%\dedication{}

% Affiliations: Please provide academic titles (Prof. or Dr.) for all authors where applicable, and include an institutional email address for all corresponding authors
%\begin{affiliations}
%Kai Xu, Luis P\'{e}rez-Fidalgo, Dr. M.~Isabel Alonso, Prof. Alejandro R.~Go\~ni\\

\noindent 
$^1$Institut de Ci\`encia de Materials de Barcelona, ICMAB-CSIC, Campus UAB, 08193 Bellaterra, Spain
% wellerm1@cardiff.ac.uk;b.charles@bristol.ac.uk;kaixu@icmab.es;lperez2@icmab.es;isabel@icmab.es

%Dr. Bethan L. Charles\\
\noindent
$^2$Dept. of Chemistry \& Centre for Sustainable Chemical Technologies, University of Bath, Claverton Down, Bath BA2 7AY, UK
%https://orcid.org/0000-0002-7876-7899

%Dr. Bethan L. Charles\\
\noindent
$^3$Dept. of Mechanical Engineering, Queens Building, University of Bristol, Bristol BS8 1TR, UK

%Prof. Mark T.~Weller\\
\noindent
$^4$Dept. of Chemistry \& Centre for Sustainable Chemical Technologies, University of Bath, Claverton Down, Bath BA2 7AY, UK

%Prof. Mark T.~Weller\\
\noindent
$^5$Dept. of Chemistry, Cardiff University, Wales CF10 3AT, UK

%Prof. Alejandro R.~Go\~ni\\
\noindent
$^6$ICREA, Passeig Llu\'is Companys 23, 08010 Barcelona, Spain\\
*Email: goni@icmab.es
%\end{affiliations}
\vspace{1cm}

% Keywords: Please provide a minimum of three and a maximum of seven keywords, separated by commas


\begin{abstract}

The exceptional optoelectronic properties of metal halide perovskites (MHPs) are presumed to arise, at least in part, from the peculiar interplay between the inorganic metal-halide sublattice and the atomic or molecular cations enclosed in the cage voids. The latter can exhibit a roto-translative dynamics, which is shown here to be at the origin of the structural behavior of MHPs as a function of temperature, pressure and composition. The application of high hydrostatic pressure allows for unraveling the nature of the interaction between both sublattices, characterized by the simultaneous action of hydrogen bonding and steric hindrance. In particular, we find that under the conditions of unleashed cation dynamics, the key factor that determines the structural stability of MHPs is the repulsive steric interaction rather than hydrogen bonding. Taking as example the results from pressure and temperature-dependent photoluminescence and Raman experiments on MAPbBr$_3$ but also considering the pertinent MHP literature, we provide a general picture about the relationship between the crystal structure and the presence or absence of cationic dynamic disorder. The reason for the structural sequences observed in MHPs with increasing temperature, pressure, A-site cation size or decreasing halide ionic radius is found principally in the strengthening of the dynamic steric interaction with the increase of the dynamic disorder. In this way, we have deepened our fundamental understanding of MHPs; knowledge that could be coined to improve performance in future optoelectronic devices based on this promising class of semiconductors.

\end{abstract}

\section{Introduction}

Metal halide perovskites (MHPs) are nowadays the focus of intense fundamental as well as applied research mainly for their exceptional photovoltaic properties that have catapulted solar cell efficiencies to values in excess of 25\%\cite{NREL} but using low-cost, solution-processing methods. MHPs with general formula ABX$_3$, being B a metal (Pb or Sn) and X a halogen atom (Cl, Br, I), are characterized by a labile inorganic cage of corner-sharing BX$_6$ octahedrons, enclosing the loosely bound atomic or molecular A-site cations in their voids. According to Goldschmidt's tolerance-factor criterium,\cite{golds26a} the A-site cations fitting in the inorganic cage voids are Cs and organic molecules such as methylammonium (MA) or formamidinium (FA). Because the A-site cations are only loosely bound to the inorganic cage by electrostatic forces, they are able to freely move (translate, rotate and librate) inside the cage voids. It is an experimentally and theoretically well-established fact that in cubic and tetragonal phases of MHPs, such dynamics is fully or partially (in-plane) unfolded, respectively, whereas in less symmetric orthorhombic phases the A-site cations are locked in certain positions and orientations inside the voids.\cite{frost16a} For example, experimentally the MA and/or FA dynamics has been directly assessed by ultra-fast vibrational spectroscopy\cite{bakul15a,selig17a} or indirectly inferred from the analysis of the atomic displacement parameter in neutron scattering\cite{welle15a,weber18a} and X-ray diffraction experiments.\cite{szafr16a} In the case of the MA$^+$ ions in pure lead halide perovskites, the dynamics consists essentially of a fast (ca. 0.3 ps) wobbling-in-a-cone motion and much slower, jump-like reorientation rotations of the molecules by 90$^\circ$.\cite{selig17a} The latter, which are the main cause of dynamic disorder, exhibit characteristic jump times ranging from 1 to 3 ps, depending on the halide atom. However, in mixed-halide compounds, these times can be as long as 15 ps.\cite{selig17a} Theoretically, the A-site cation dynamics has been well accounted for within molecular-dynamics calculations.\cite{ghosh17a,ghosh19a,maity23a} Using a diffusive model,\cite{matto15a} ab-initio molecular dynamics simulations yield for MAPbBr$_3$ at 300 K\cite{maity23a} a relaxation time of ca. 340 ps for the fast motion and about 2 ps for the jump-like rotations, in excellent agreement with the experiment. This dynamics has direct impact on one of the distinctive features of MHPs, namely the interplay between the inorganic network and the atomic or molecular cations enclosed in the cage voids, determining, at least in part, the outstanding optoelectronic properties of these semiconductor materials.
\vspace{0.25cm}

The interplay between the inorganic metal-halide sublattice and the network of A-site cations picks up contributions from two interactions with different origin and acting at different length scales: Hydrogen bonding and steric effects. Hydrogen bonding results from the electrostatic interaction between the hydrogen atoms of the organic cations and the negatively charged halide anions. %\cite{H-bond}
In the case of the Cs$^+$ cations, H bonding is replaced by the bare electrostatic anion-cation attraction. In contrast, steric effects corresponds to non-bonding dipole-dipole interactions between molecules and/or atoms, which are well described by a Lennard-Jones potential. %\cite{steric}
At large distances steric effects correspond to the weak van der Waals attraction that is much weaker than electrostatic interactions, being thus negligible against H bonding.  However, at short distances the repulsion between the electronic clouds of neighboring atoms or molecules comes into play and the steric interaction becomes strongly repulsive. In the case of MHPs, steric effects are intimately related to the movement of the A-site cations inside the cage voids, which provides the necessary kinetic energy to bring cations and anions sufficiently close together. Hence, at the risk of being redundant, the steric repulsion will be hereafter called dynamic steric interaction (DSI). H bonding is ubiquitous in hybrid halide perovskites and has been repeatedly invoked to explain the structural phase behavior of MA lead halides as a function of temperature\cite{yinxx17a,maity23a} and pressure.\cite{capit17a,yesud20a} Apart from contributing to the structural stability of the low-temperature orthorhombic phases of MHPs, first-principle calculations have shown that H bonding is instrumental for the tilting of the PbX$_6$ octahedrons.\cite{leexx16a,leexx16b} Furthermore, molecular dynamics simulations have highlighted the role that H bonding plays at the one-to-one connection between octahedral tilting and local structural deformations with the roto-translational dynamics of the molecular cations.\cite{ghosh17a,ghosh19a,maity23a} This is the origin of the dynamic disorder caused by unleashed A-site cation dynamics. Curiously, besides for its consideration to explain phase stability in inorganic MHPs,\cite{leexx16a} the dynamic steric interaction has been widely ignored in the literature. Yet, here we will show that DSI is crucial for a final understanding of the structural phase sequences observed in MHPs as a function of pressure and halide composition.
\vspace{0.25cm}

For MHPs Raman scattering turns out to be a very powerful technique, since it grants easy access and without experimental complications to the degree of dynamic disorder present in the sample for given temperature and pressure conditions. In previous temperature-dependent experiments on the three MA lead halides, we found evidence of the coupling mentioned before between the vibrations of the anionic network PbX$_3$ (X=I, Br, Cl) and the MA cations in the Raman scattering signature.\cite{brivi15a,leguy16a} As a consequence of the steric interaction between the MA molecules and the halogen atoms of the inorganic cage and due to dynamic disorder, the vibrational modes of the cage exhibit a wide statistical distribution of frequencies, which in turn leads to a strong {\it inhomogeneous} broadening of the Raman peaks. In contrast, in the low-temperature orthorhombic phase, when the organic cations are locked and ordered inside the cage voids, becoming well oriented along high-symmetry directions of the perovskite crystal, dynamic disorder just disappears. The result is a pronounced reduction of the linewidths of the Raman peaks, which is readily observed in low-temperature Raman spectra.\cite{yinxx17a,brivi15a,leguy16a,sharm20a} Interestingly, a similar locking effect of the MA cations and the concomitant reduction in linewidth of the inorganic cage phonons can be induced at room temperature through the application of high hydrostatic pressure.\cite{ghosh19a,capit17a,franc18a} Here we make explicit use of this spectroscopic tool to monitor the appearance or disappearance of structural disorder as a function of pressure and temperature in relation to the A-site cation dynamics.
\vspace{0.25cm}

\begin{figure}[ht]
  \includegraphics[width=6.5cm]{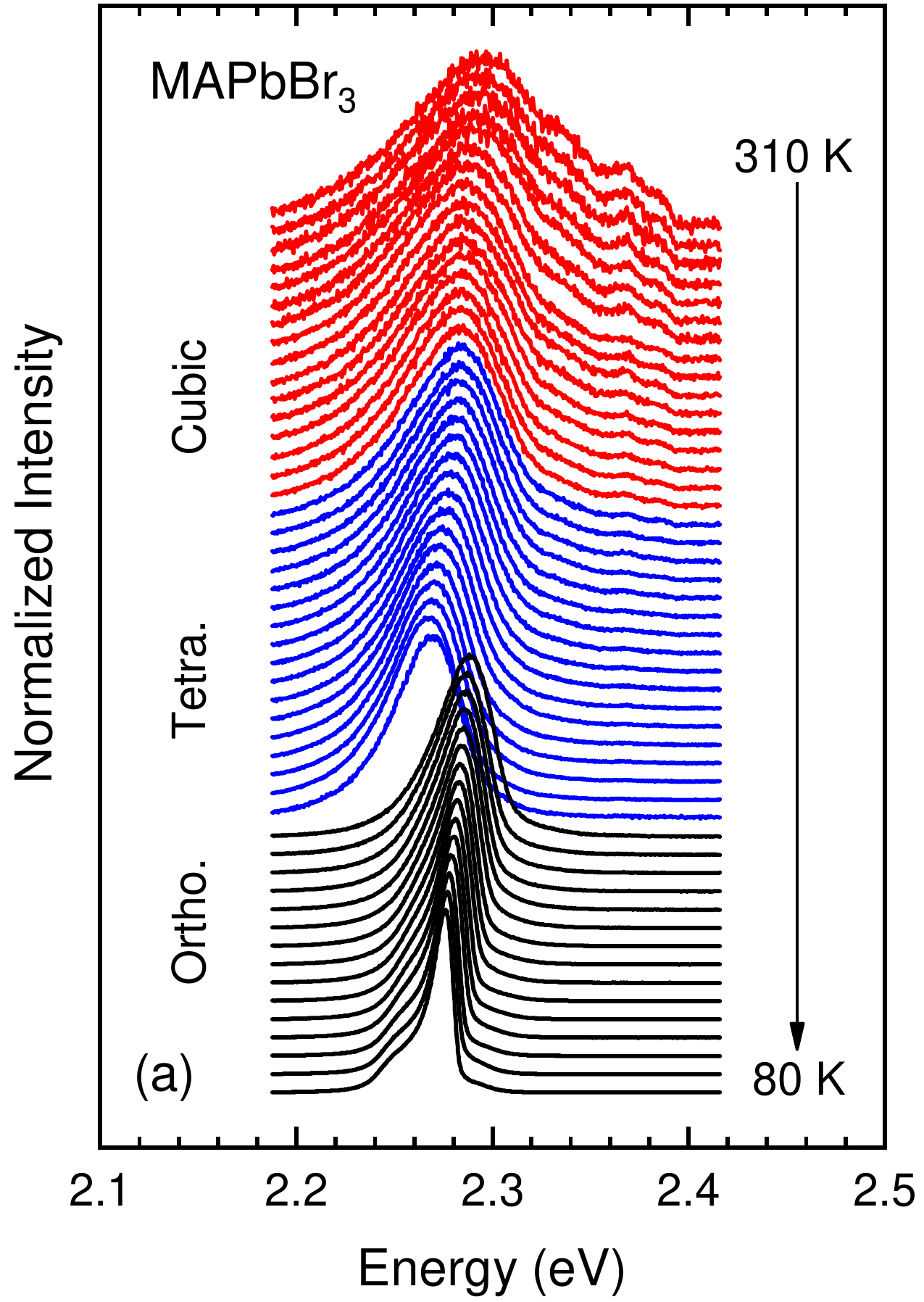}
  \includegraphics[width=6.5cm]{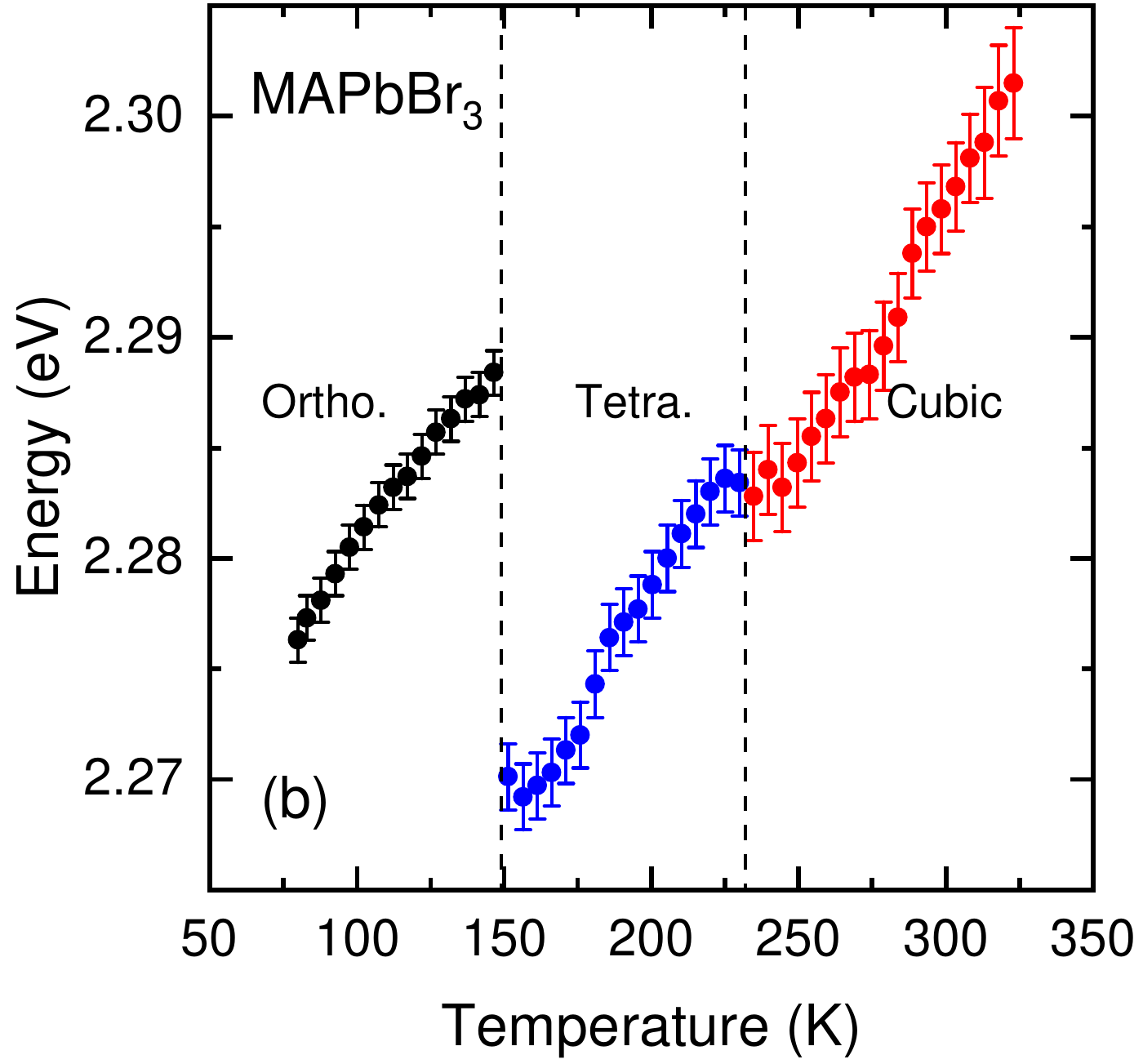}
  \caption{(a) PL spectra of a MAPbBr$_3$ single crystal recorded at different temperatures using the green laser line (514.5 nm) for excitation. The spectra were normalized to their maximum intensity and plotted with a vertical shift for clarity. The temperature range is indicated (temperature step ca. 5 K). The different colors represent the different structural phases adopted as a function of temperature (Cubic: Pm$\overline{3}$m, Tetra.: I4/mcm, Ortho.: Pnma). (b) The maximum PL peak energy position plotted as a function of temperature, obtained from the PL lineshape fits to the spectra shown in (a), using a cross-product function (Eq. (1) of Supporting Information). Dashed lines indicate the phase transition temperatures.}
  \label{PL vs T}
\end{figure}

In this work, we present a systematic study of the structural phase behavior of high-quality MAPbBr$_3$ single crystals as a function of temperature in the range 80 to 320 K and ambient pressure as well as a function of pressure up to ca. 7 GPa at room temperature. This has been accomplished by monitoring the temperature and pressure-induced changes in the fundamental band gap and vibrational spectrum of MAPbBr$_3$, as observed in PL and Raman experiments, respectively, following the procedure reported elsewhere.\cite{franc18a,franc20a} By combining the results obtained here for MAPbBr$_3$ with data from the available literature on temperature and/or pressure-dependent studies for MAPbI$_3$,\cite{welle15a,szafr16a,leguy16a,franc18a,pogli87a,onoda90a,onoda92a,capit16a,wangx17a,jaffe16a,ouxxx16a,jiang16a,kongx16a} MAPbBr$_3$,\cite{maity23a,yinxx17a,capit17a,yesud20a,leguy16a,pogli87a,onoda90a,onoda92a,jaffe16a,kongx16a,swain07a,manni20a,wangx15a} MAPbCl$_3$,\cite{leguy16a,pogli87a,onoda90a,onoda92a,wangx16b} FAPbI$_3$,\cite{carpe23a} FAPbBr$_3$,\cite{manni20a,wangx16a} FA$_x$MA$_{1-x}$PbI$_3$,\cite{franc20a,mohan19a,sharm21a} CsPbI$_3$,\cite{yixxx22a} CsPbBr$_3$,\cite{manni20a} and the data from two recent reviews,\cite{lixxx20a,celes22a} we were able to conceive a general picture about the relationship between crystal structure and dynamic disorder in MHPs. One particularly important finding is that at the temperatures for unfolded A-site cation dynamics, the structural stability of the crystalline phases observed for increasing pressure, A-site cation size or decreasing halogen atomic radius can only be understood in terms of a strengthening of the DSI, rather than due to H-bonding effects. Moreover, we offer an explanation for the fact that the onset of the pressure-induced amorphization, i.e. static disorder, based on the amounts of vacancies present in each particular sample. We note that we have intentionally excluded the results on thin films and nanocrystals from the discussion to avoid complications due to effects on structural behavior of grain boundaries, interfaces, surfaces, and/or confinement, making the underlying physics difficult to understand.
\vspace{0.25cm}

\section{Results}

\subsection{Temperature and Pressure-Dependent Photoluminescence (PL) Spectra}

Figure \ref{PL vs T}a shows the evolution with temperature of the PL spectra of a MAPbBr$_3$ high-quality single crystal in the range from 310 to 80 K. All spectra were normalized to their absolute maximum intensity and vertically offset to ease their comparison. The different colors correspond to the temperature ranges of stability of the different crystalline phases of MAPbBr$_3$, as indicated. According to X-ray diffraction results,\cite{pogli87a,swain07a,yinxx17a} starting at ambient, the phase sequence for decreasing temperature is $\alpha$-cubic $\rightarrow$ $\beta$-tetragonal-I $\rightarrow$ $\gamma$-tetragonal-II $\rightarrow$ $\delta$-orthorhombic. The $\gamma$ phase is not indicated in Fig. \ref{PL vs T}a because we missed it in our experiments, since it has a very narrow stability range of 5 K. At all temperatures a single peak dominates the PL spectra of MAPbBr$_3$, corresponding to the free-exciton emission.\cite{franc20a,galco17a} With decreasing temperature the exciton peak exhibits a monotonous redshift of its energy, except for the sudden jumps at the phase transitions, and a clear decrease in linewidth. In view of the relatively small binding energy of ca. 15 meV,\cite{tilch16a} the redshift can be taken as representative of the temperature dependence of the fundamental band gap. The linewidth reduction, in turn, is indicative of a homogeneously broadened emission peak, which means it is lifetime limited. In high quality crystals, non-radiative exciton decay is mainly associated to the scattering by phonons, thus, being strongly temperature dependent. At low temperatures below ca. 110 K, several peaks become apparent at the low-energy side of the main exciton peak (see Fig. \ref{PL vs T}a), which are ascribed to emission from bound (acceptor/donor) exciton complexes, as reported elsewhere.\cite{franc21a}
\vspace{0.25cm}

To analyze the PL spectra of the hybrid perovskites we used a Gaussian-Lorentzian cross-product function for describing the main emission peak, as successfully employed for the analysis of the PL spectra of MAPbI$_3$\cite{franc18a} and MA/FA mixed crystals.\cite{franc20a} The expression for the cross-product function is given in the Supporting Information. It contains three adjustable parameters: The amplitude prefactor $A$, the peak energy position $E_0$, and the full width at half maximum (FWHM) $\Gamma$. This function is a useful simplification of a Voigt function, which corresponds to the mathematical convolution of a Lorentzian and a Gaussian. There is an additional lineshape parameter which takes the values $s=0$ for pure Gaussian and $s=1$ for pure Lorentzian. For MAPbBr$_3$ the exciton emission lineshape turned out mainly Gaussian with little Lorentzian admixture. The values of the peak energy $E_0$ are plotted as a function of temperature in Fig. \ref{PL vs T}b (the PL linewidths and intensities are shown in Fig. S1 of the Supporting Information). As mentioned above, we consider the shift of the PL peak energy $E_0$ with temperature representative of the temperature change of the gap. The linear increase of the gap with increasing temperature observed for the cubic and tetragonal phases is a common trend of MHPs, which was explained as due to two equally-contributing effects, namely thermal expansion and enhanced electron-phonon interaction.\cite{franc19a} Furthermore, the temperatures at which the jumps in the gap energy occur are in excellent agreement with the phase transition temperatures from X-ray data (dashed lines in Fig. \ref{PL vs T}b). At the phase transitions, the gap always increases for the phase with lower symmetry. This is due to the sudden increment in the overall octahedral tilting of the PbBr$_6$ octahedrons, which leads to a strong reduction of the Pb-Br-Pb bond angle, reducing the overlap between valence Pb and Br orbitals and increasing the bandgap.\cite{jaffe16a,kongx16a}
\vspace{0.25cm}

\begin{figure}[ht]
  \includegraphics[width=6.5cm]{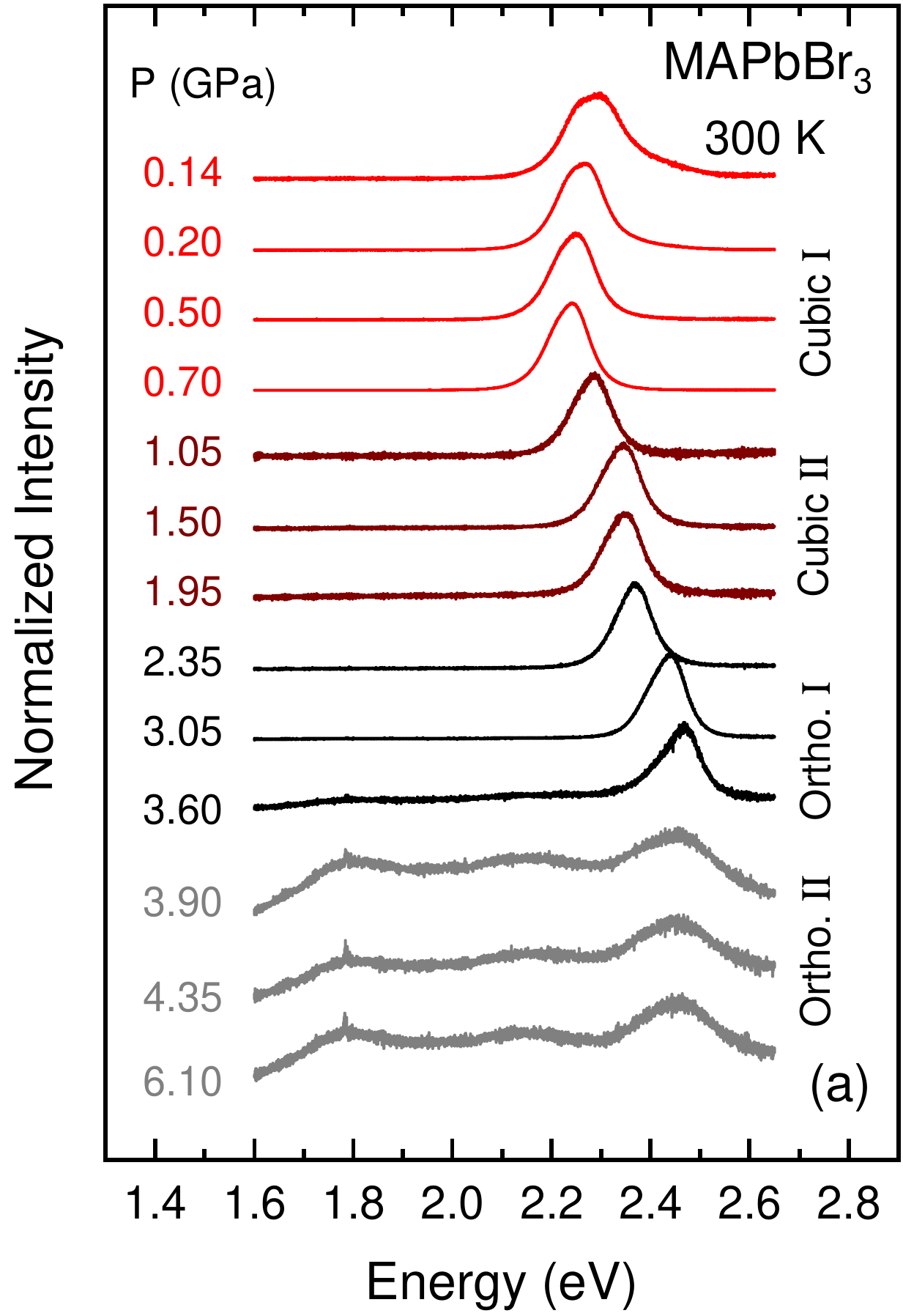}
  \includegraphics[width=6.5cm]{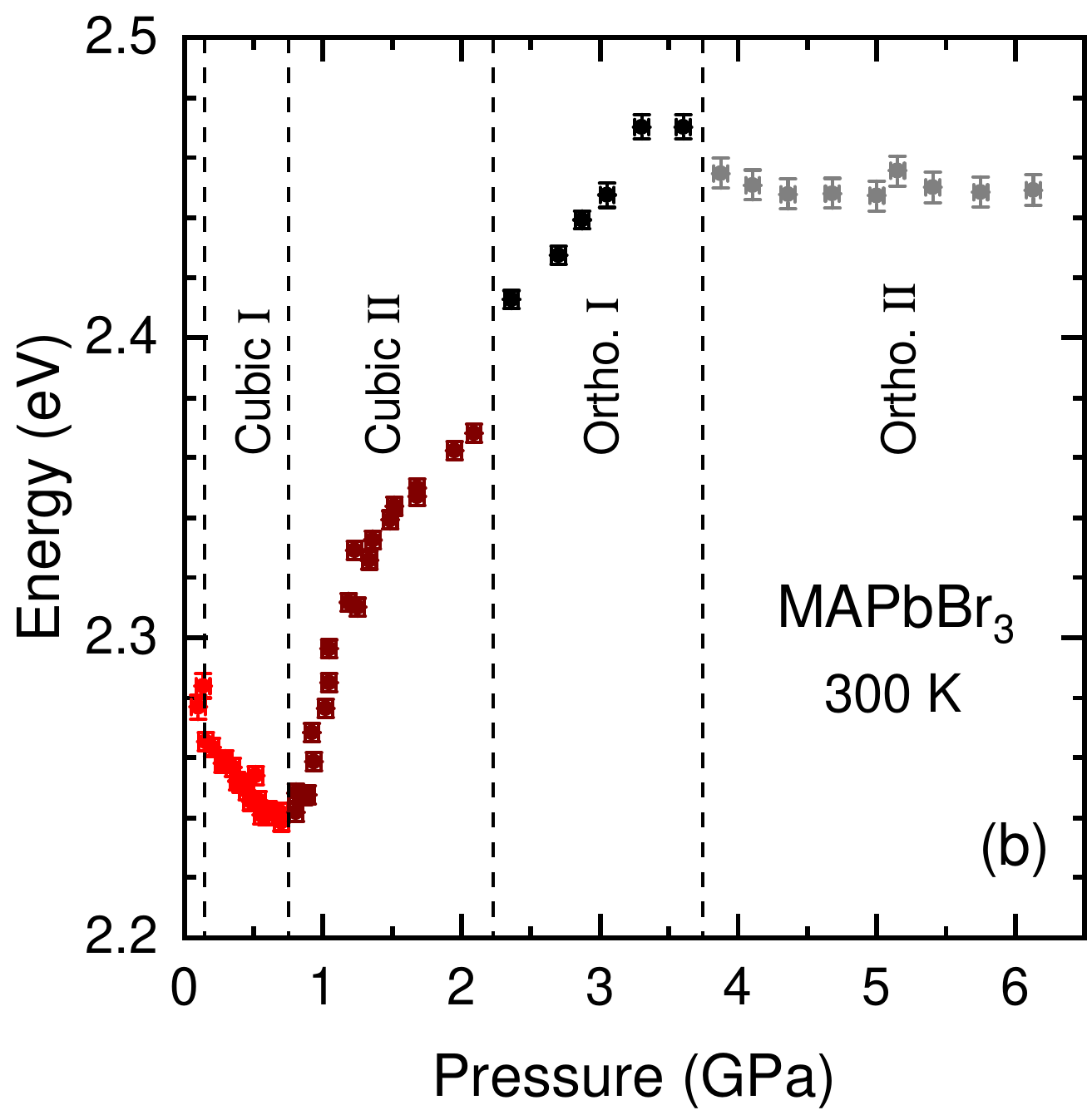}
  \caption{(a) PL spectra of MAPbBr$_3$ obtained for incremental steps of pressure up to about 6.5 GPa using the 405-nm laser line for excitation. The spectra were normalized to their maximum intensity and plotted with a vertical shift for increasing pressure. The different colors indicate the subsequent phases adopted by the material during the pressure upstroke (Cubic I: Pm$\overline{3}$m, Cubic II: Im$\overline{3}$, Ortho. I: Pnma, Ortho. II: unknown). (b) The PL peak energy $E_0$ plotted as a function of pressure, obtained from the PL lineshape fits using Eq. (1) of the Supporting Information. The pressures at which the phase transitions occur are indicated by vertical dashed lines. See text for details.}
  \label{PL vs P}
\end{figure}

Figure \ref{PL vs P}a shows representative PL spectra of MAPbBr$_3$ measured at different pressures up to about 6.5 GPa. Spectra were again normalized to its absolute maximum intensity and vertically offset to ease their comparison. The main PL peak exhibits abrupt changes in the position of its maximum, which are indicative of the occurrence of three phase transitions in the pressure range of the experiment. This can be better appreciated in Fig. \ref{PL vs P}b, where the values of the peak energy $E_0$ are plotted as a function of pressure. Different colors correspond to the four observed phases and vertical dashed lines (except for the first one) mark the corresponding phase transition pressures. We note that at the very beginning solely of the first pressure upstroke, a sudden redshift of the PL emission, i.e. of $E_0$, is observed to occur in the range from 0 to $\sim0.25$ GPa. As shown below, this effect is accompanied by changes in the linewidth of the Raman peaks. Since this happens only once, we believe it is not related to a phase transition. We speculate that this behavior might arise from an initial strain relaxation the first time the sample is pressurized in the diamond anvil cell (DAC). Such a strain could have been introduced by the way the small chips to be loaded into the DAC are produced (see Methods section).
\vspace{0.25cm}

Within the cubic-I (Pm$\overline{3}$m) phase, stable from ambient conditions, the PL spectra exhibit a clear redshift and the gap energy of MAPbBr$_3$ displays a negative linear dependence on pressure. A linear regression to the data points yields a pressure coefficient of ($-54\pm5$) meV/GPa, which is very similar to that of other counterparts like MAPbI$_3$\cite{franc19a} and MAPbCl$_3$\cite{wangx16b} (for comparison see the survey of pressure coefficients of MHPs published in Ref.\cite{franc19a}). As previously argued for MAPbI$_3$\cite{franc18a}, such a negative pressure dependence of the gap can be readily explained using the well-established systematic about the pressure coefficients of conventional semiconductors\cite{gonix98a} and accounting for the bonding/antibonding and atomic orbital character of the valence and conduction-band states. Relativistic band-structure calculations\cite{frost14a,evenx15a} for a pseudo-cubic phase of MAPbI$_3$ predict that due to the huge spin-orbit interaction present in heavy atoms like Pb, there is a so-called band inversion. For MHPs this means that the top of the valence band is predominantly composed by antibonding Pb $6s$ orbitals, which shift up in energy with pressure, whereas the bottom of the conduction band is formed by the antibonding split-off Pb $6p$-orbitals, which are fairly pressure insensitive. A totally similar result is expected for MAPbBr$_3$, which explains the negative sign and magnitude of the gap pressure coefficient.
\vspace{0.25cm}

In MAPbBr$_3$ the first phase transition thus occurs at a low pressure of ca. 0.75 GPa, as signalled by a turnover in the change of the PL peak energy $E_0$ with pressure. This is an isostructural transition because the new high-pressure phase, which is stable up to 2.2 GPa, corresponds to the cubic-II (Im$\overline{3}$) phase, as reported elsewhere.\cite{capit17a,yesud20a,jaffe16a} This phase is characterized by a stepwise linear increase of $E_0$ with increasing pressure, exhibiting a kink in the pressure dependence at about 1.2 GPa, also observed by Yesudhas et al.\cite{yesud20a} However, there is no hint to a phase transition from the Raman data at this pressure, thus, the reason for the kink remains elusive to us. In contrast, the fact that in the cubic-II phase the gap energy gradually but steadily increases with pressure can be understood as arising from a pressure-induced increase in octahedral tilting. The cubic-II phase (Im$\overline{3}$) is obtained from the cubic-I (Pm$\overline{3}$m) by an alternate tilting of the PbBr$_6$ octahedrons in the direction of the cube diagonals. This doubles the unit cell in all three directions that remains cubic. Once the tilting starts, it increases gradually with pressure, causing the observed incremental opening of the gap.
\vspace{0.25cm}

The second phase transformation occurs at 2.2 GPa and is characterized by an abrupt increase of the PL peak energy. As explained in the discussion of the Raman results, this new phase is perfectly crystalline and probably orthorhombic in nature, in agreement with previous reports.\cite{capit17a,yesud20a} The PL spectra clearly indicate the occurrence of a third phase transition at about 3.75 GPa, characterized by a dramatic change in PL lineshape (gray spectra in Fig. \ref{PL vs P}a). Two broad peaks appear at much lower energies and there is an overall lost of intensity together with a pronounced broadening of the main peak (see Fig. S2 of the Supporting Information). This speaks for a large heterogeneity in the sample, as far as the electronic states involved in the optical transitions are concerned. In fact, this is the pressure range for which a  a pressure-induced amorphization is reported for MAPbBr$_3$.\cite{capit17a,yesud20a,jaffe16a,wangx15a} However, we anticipate that the Raman data indicate again that this phase is crystalline and probably orthorhombic up to the highest pressure of this experiment close to 6.5 GPa. We will return to deal with the amorphization and how it might be generated under pressure in the discussion of the Raman results. Finally, we remark that the changes in the PL emission (the Raman too) are fully reversible only provided the pressure was kept below that of the transition into the ortho-I phase. Otherwise there is a certain degree of hysteresis in the PL peak energy by releasing the pressure.
\vspace{0.25cm}

\subsection{Temperature and Pressure-Dependent Raman Spectra}

\begin{figure}[ht]
  \includegraphics[width=6.5cm]{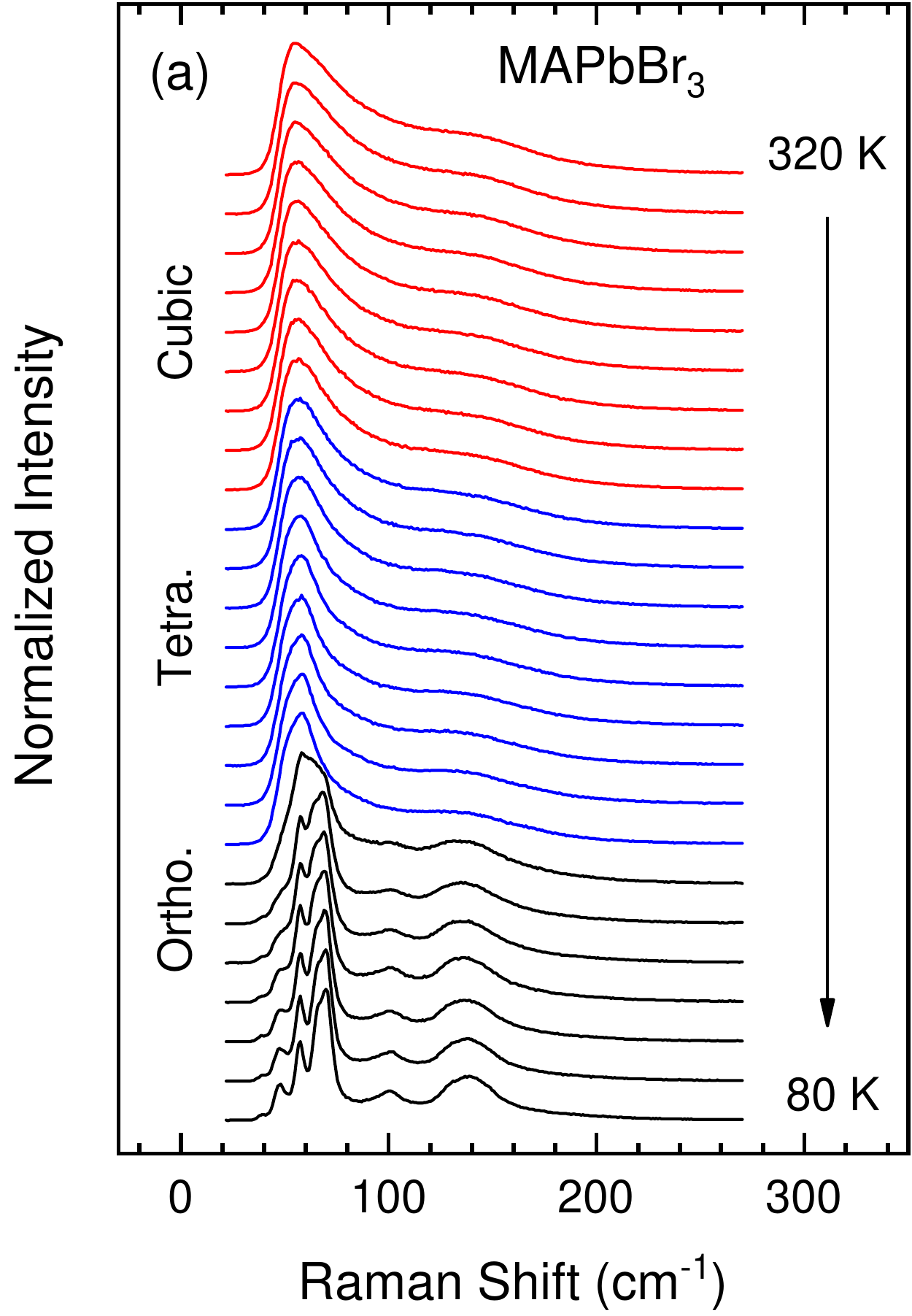}
  \includegraphics[width=6.5cm]{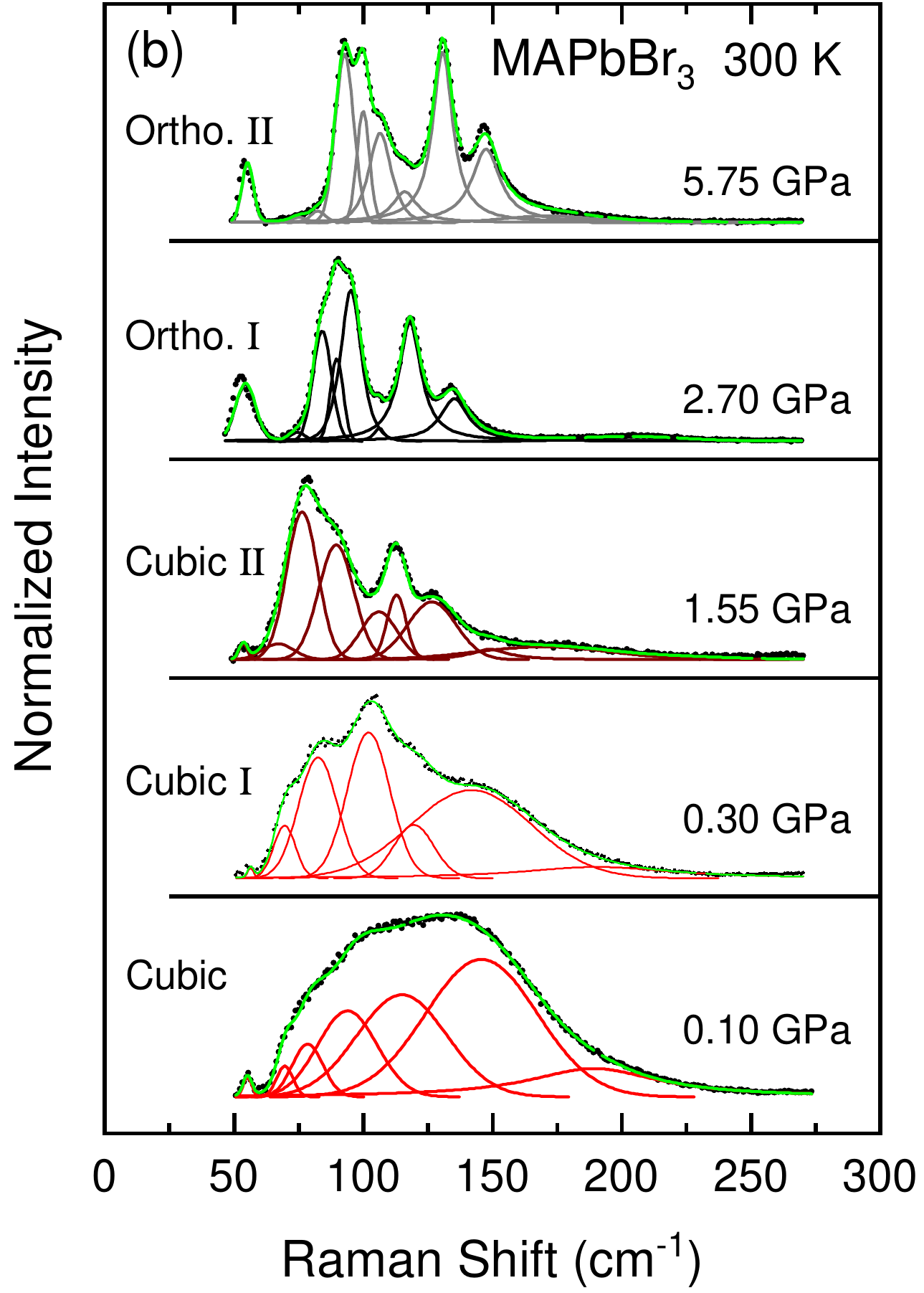}
  \caption{(a) Raman spectra of MAPbBr$_3$ measured at different temperatures from 320 K to 80 K (steps of 10 K) in the spectral range of the inorganic cage phonon modes using the 785-nm line for excitation. The spectra were normalized to their maximum intensity and shifted vertically for clarity. The different colors of the spectra indicate the changes in phase after every transition. (b) Examples of the performed lineshape fits (green solid curves) to the Raman spectra (black closed symbols) using Gaussian functions for different pressures, as indicated, each one representing a different high-pressure phase of the material. The solid curves in the color of the corresponding phase represent the different phonon components. The shown spectra were obtained by subtracting a special function used for describing the combined effect of the dichroic filter and the broad central peak (see text for details).}
  \label{Raman spectra}
\end{figure}

Figure \ref{Raman spectra}a summarizes the Raman results obtained on single-crystalline MAPbBr$_3$ as a function of temperature in a similar range as for the PL measurements (80 - 320 K) using the 785-nm line for excitation. The Raman spectra shown here correspond to the spectral region of the inorganic cage phonon modes below 300 cm$^{-1}$.\cite{brivi15a}  As reported before,\cite{leguy16a} in the high-temperature cubic phase (red spectra), the MA dynamics is fully unfolded, resulting in a strong inhomogeneous broadening of the inorganic cage phonons due to the strong coupling to the molecular cations. In fact, the Raman spectra are quite featureless, exhibiting essentially a broad peak centered at around 70 cm$^{-1}$. The width of this Raman band decreases slightly, when the sample transforms into the tetragonal phase (blue spectra), for which the MA cations are free to move only in the tetragonal plane. The partial reduction of the dynamic disorder in the tetragonal phase leads to a slight decrease of the inhomogeneous broadening. In stark contrast, several well-defined peaks are apparent in the Raman spectra of the orthorhombic phase (black curves in Fig. \ref{Raman spectra}a), a phase in which the MA cations are locked inside the cage voids in a state of static order. Concomitant with the disappearance of dynamic disorder, the inhomogeneous broadening vanishes, such that the Raman peaks just display their lifetime-limited homogeneous linewidth. An instructive digression on the relative importance of homogeneous versus inhomogeneous broadening in the Raman spectra of the three halide compounds MAPbX$_3$ with X=Cl, Br and I in relation to dynamic disorder is given in the Supporting Information.
\vspace{0.25cm}

The raw Raman spectra recorded for MAPbBr$_3$ under pressure are shown in Fig. S3 of the Supporting Information. For a quantitative assessment of the effect of pressure on the vibrational spectrum of MAPbBr$_3$ we have decomposed each Raman spectrum in its different mode components by a lineshape analysis, as illustrated in Fig. \ref{Raman spectra}b, where a representative example of the fits for each of the observed phases is displayed. We note that at room temperature and mainly for the first two phases (P$<2.2$ GPa), all Raman lineshapes are affected by the presence of a very broad and intense peak at very small Raman shifts and the edge-like attenuation caused by the dichroic filter used to screen the laser. The former is interpreted as a broad central Raman peak originating from local polar fluctuations in the perovskite structure caused by dynamic disorder \cite{yaffe17a}. A special function was constructed to describe such background,\cite{franc20a} which has been subtracted from the Raman spectra for a better visualization of the phonon modes. This simplifies the fitting of the Raman spectra using Gaussian functions, as illustrated in Fig. \ref{Raman spectra}b. The number of Gaussian peaks and the approximate frequency positions are consistent with previous full assignments of Raman \cite{brivi15a,leguy16a} and far-infrared spectra \cite{sendn16a} as well as those observed at low temperature for the orthorhombic phase (Fig. \ref{Raman spectra}a).
\vspace{0.25cm}

\begin{figure}[ht]
  \includegraphics[width=6.5cm]{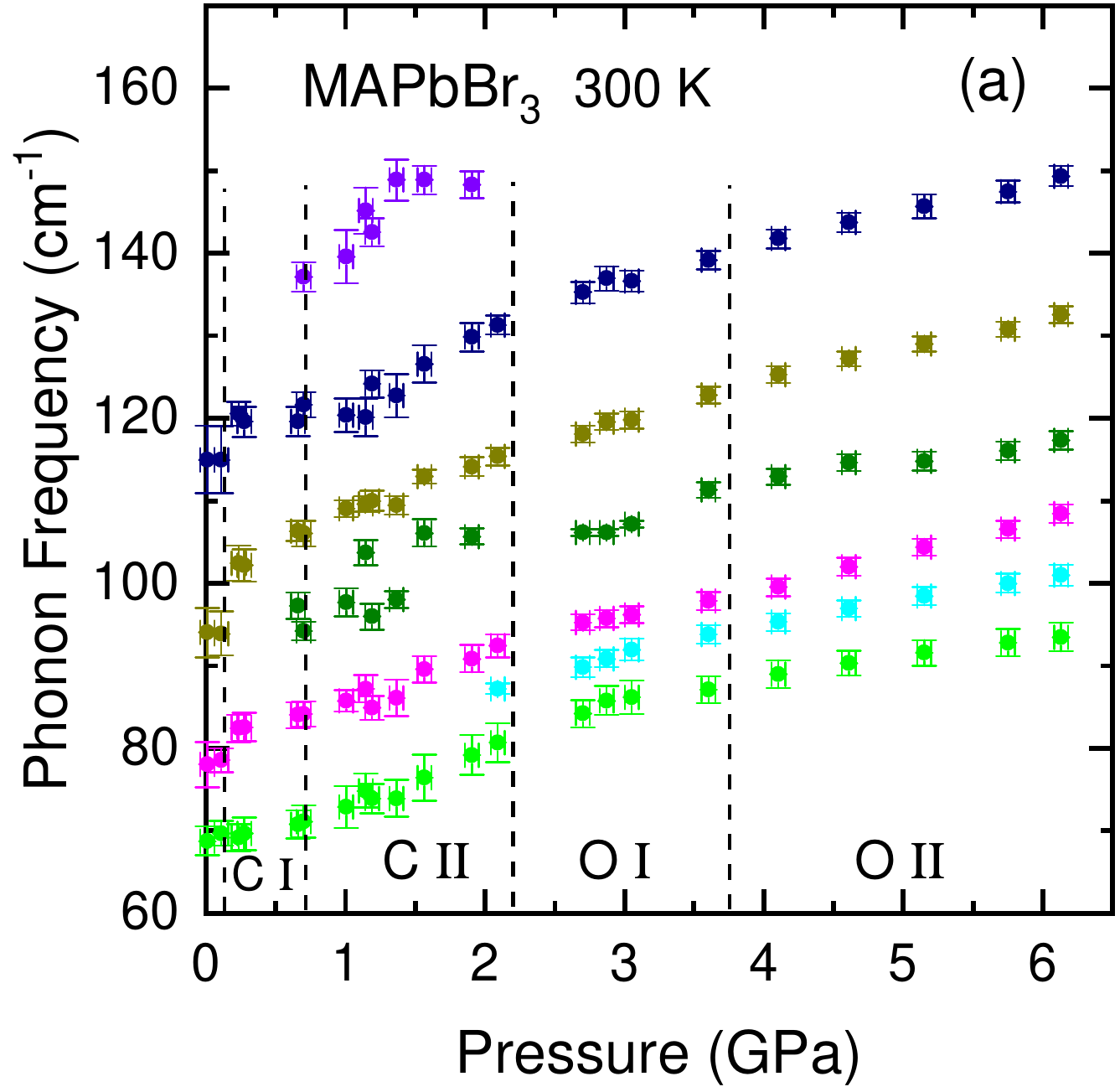}
  \includegraphics[width=6.5cm]{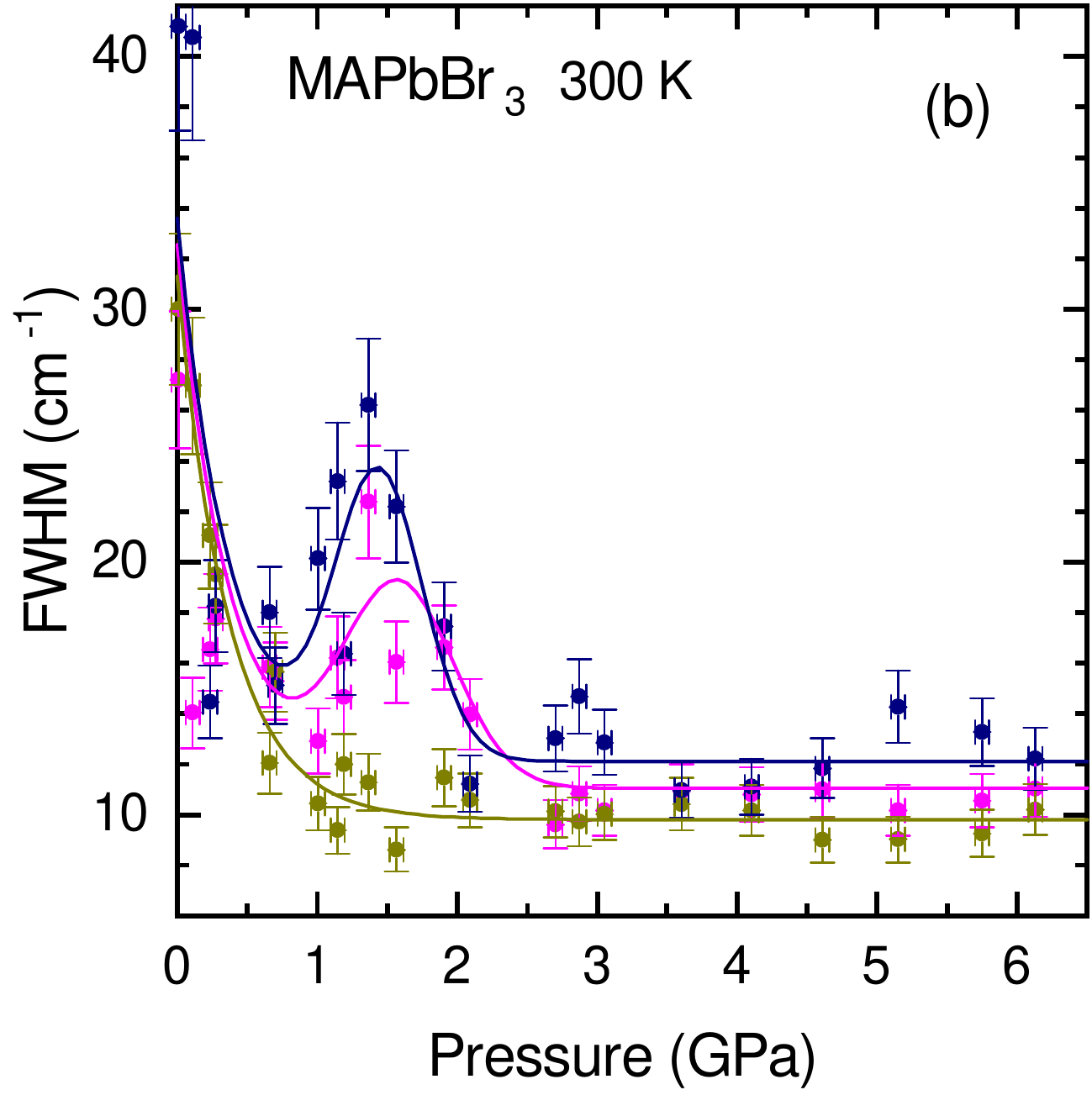}
  \caption{(a) The frequency and (b) full width at half maximum (FWHM) of the Raman peaks in the spectral region of the inorganic cage phonons below 300 cm$^{-1}$, as obtained from the lineshape fits as a function of pressure up to ca. 6.5 GPa. The pressures at which the phase transitions occur are marked with vertical dashed lines. The different phases are indicated (C: cubic, O: orthorhombic). %Representative error bars are indicated by the symbols inside the dashed squares.
  }
  \label{Raman param}
\end{figure}

The results of the Raman lineshape fits for the frequency and FWHM of the main peaks apparent in the Raman spectra of MAPbBr$_3$ (see Figs. S3 and \ref{Raman spectra}b) are depicted in Fig. \ref{Raman param} as a function of pressure. Unlike what happened with MAPbI$_3$\cite{franc18a}, the changes in the Raman frequencies and mainly the linewidths were much less pronounced in MAPbBr$_3$. The latter is a consequence of the greater {\it homogeneous broadening} of the Raman peaks due to a stronger coupling with the MA molecules within the narrower voids of the lead bromide cage (see digression in the Supporting Information), that hampers the observation of the variations of the inhomogeneous part of the linewidth with the amount of disorder. However, the clear changes in Raman lineshape observed with increasing pressure like the reduction of the linewidths and/or the appearance of additional well-resolved peaks, allowed us to corroborate the occurrence of the phase transitions previously ascertained from the PL experiments. Therefore, the dashed lines in Fig. \ref{Raman param}a also denote the phase transition pressures. An exception is the first abrupt and irreversible decrease in linewidth, which occurs only once at the start of the pressure experiments.
\vspace{0.25cm}

An important result of this work concerns the observation of a sharpening of the Raman modes also for the orthorhombic-II phase (see Figs. \ref{Raman spectra}b and \ref{Raman param}b), a phase that appears to be stable above the onset of amorphization, according to most of the high-pressure work on MAPbBr$_3$ \cite{capit17a,yesud20a,jaffe16a,wangx15a} and other halide perovskites. \cite{szafr16a,capit16a,ouxxx16a,jiang16a,kongx16a,wangx16b} On the contrary, our Raman linewidths remain narrow up to ca. 6.5 GPa, the highest pressure of these experiments, exactly as was previously reported by us for MAPbI$_3$ too. \cite{franc18a} In this respect, it is interesting to compare our Raman results with those of Capitani {\it et al.}, \cite{capit17a} where the low-frequency Raman spectra exhibit a broad, featureless band for the cubic low-pressure phases and well-defined, sharp peaks for the orthorhombic-I phase, exactly like us. The key difference lies in the strong broadening that Capitani {\it et al.} observe in MAPbBr$_3$ above ca. 4 GPa, which was ascribed to an amorphous-like state of {\it static} disorder. \cite{capit17a,celes22a} Considering that the pressure-induced changes in the Raman linewidth are due to variations of its inhomogeneous part, this provides a tool to monitor the degree of disorder of the crystal lattice. Moreover, we note that the inhomogeneous broadening is unable to distinguish between static or dynamic disorder. At least in this particular case, this is so because the typical duration times of the Raman scattering processes are in the sub-picosecond regime, \cite{tusch18a} i.e. much faster than the times required for the jump-like reorientations of the MA molecules causing dynamic disorder. Hence, a Raman measurement just corresponds to a sampling of $10^{12}$ to $10^{13}$ different but static MA orientational motifs per second, occurring in the sample throughout the molecular-cation dynamics. In fact, a marked increase in inhomogeneous broadening has been also observed in the Raman spectra of the low-temperature phases of FAPbBr$_3$ \cite{reuve22a} and FA$_x$MA$_{1-x}$PbI$_3$ \cite{franc20a} with $x>0.4$, which exhibit static structural disorder.
\vspace{0.25cm}

\begin{figure}[ht]
  \includegraphics[width=6.5cm]{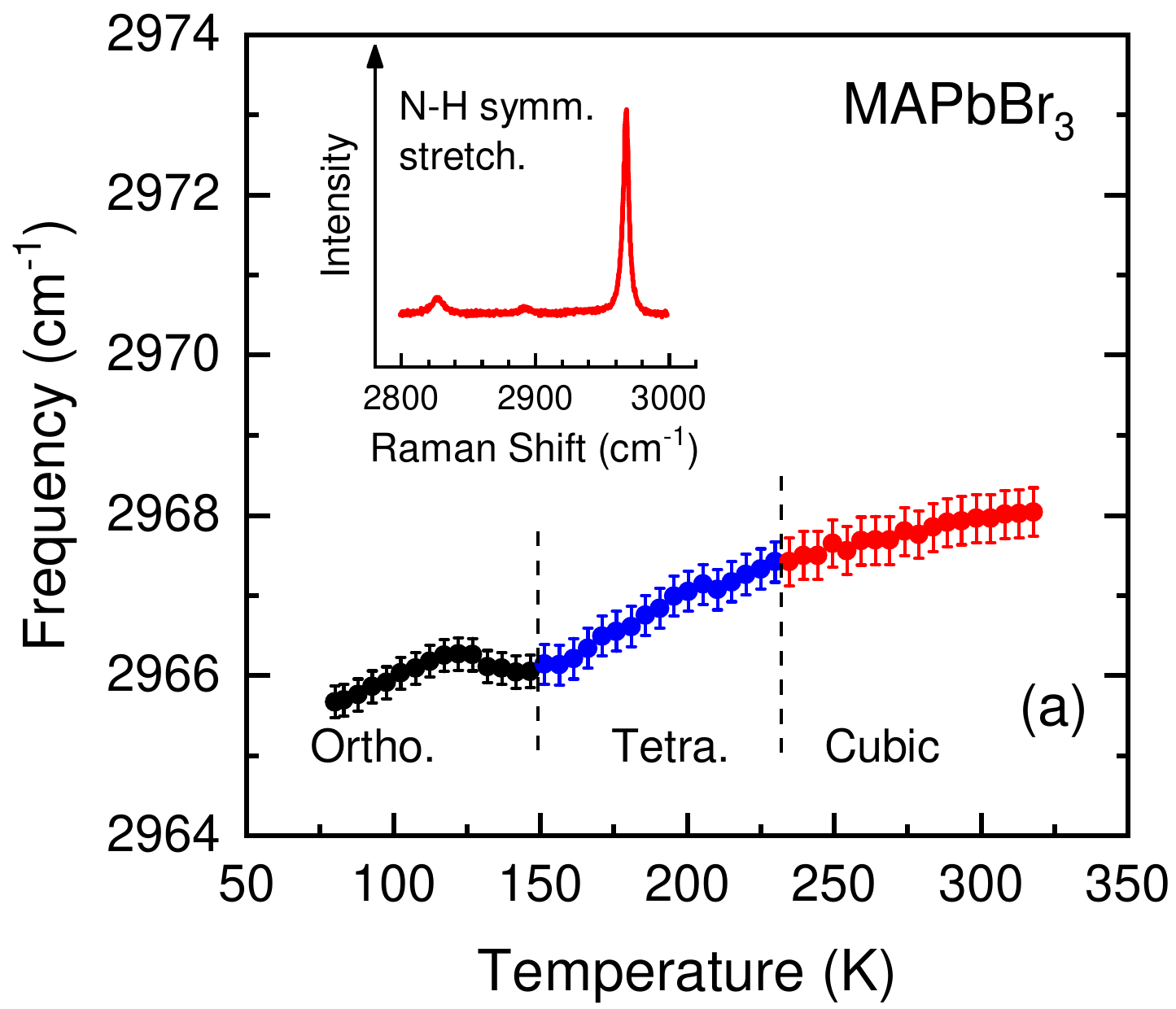}
  \includegraphics[width=6.5cm]{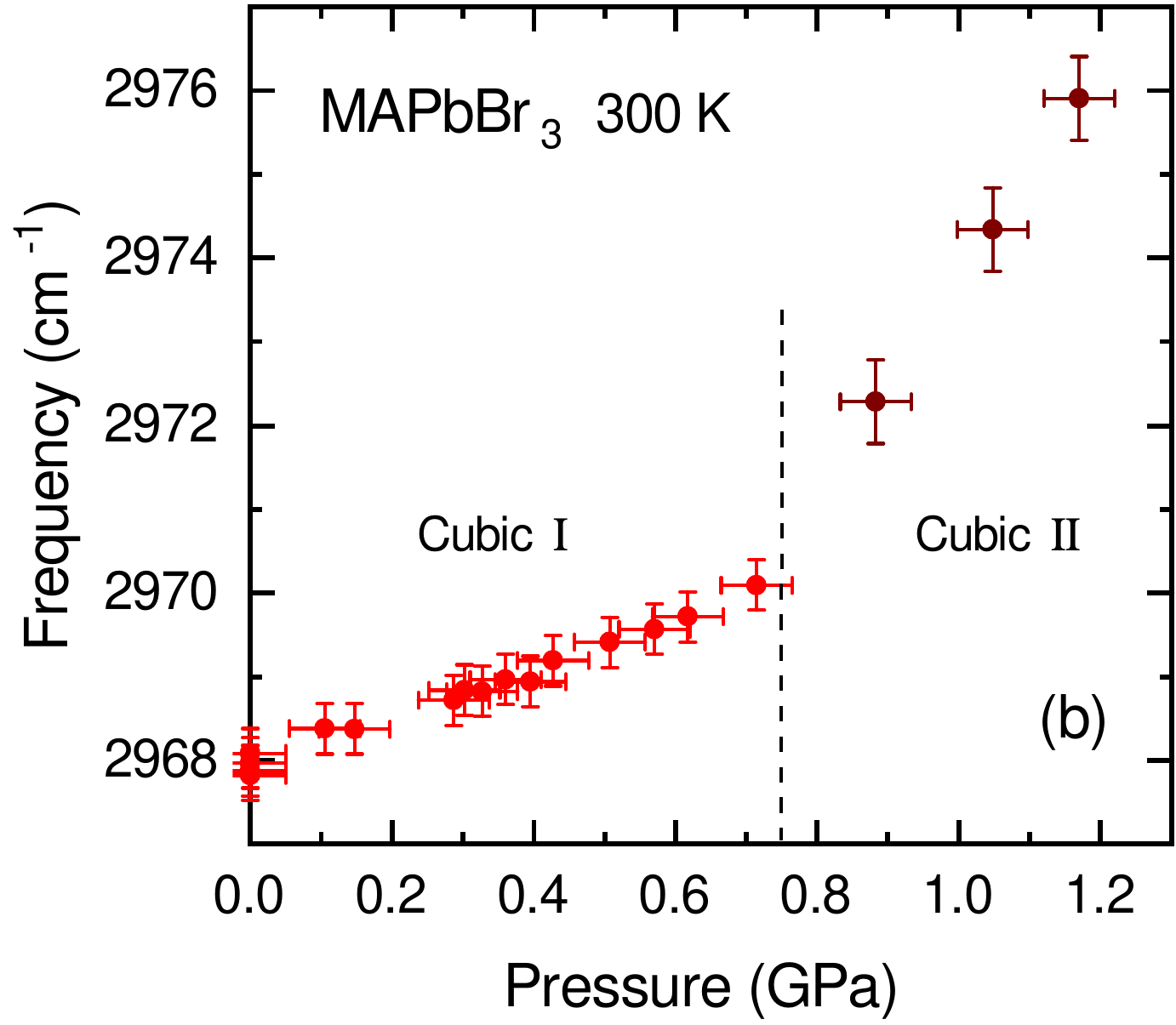}
  \caption{The frequency of the N-H symmetric stretching vibration [$\nu_s$(NH$_3^+$)] of the methylammonia (a) as a function of temperature at ambient pressure and (b) as a function of pressure at room temperature. The pressures at which the phase transitions occur are marked with vertical dashed lines and the different phases are indicated. The inset shows a representative Raman spectrum in the range of the N-H stretching vibrations around 3000 cm$^{-1}$.
  }
  \label{N-H stretch}
\end{figure}

Hence, according to the Raman data, with increasing pressure MAPbBr$_3$ transforms from a state of dynamic disorder in the cubic phases, due to an unleashed MA-cation dynamics, to a state of static order with all MA molecules locked inside the cage voids and orderly oriented in the repeated unit cell of the orthorhombic-I phase. Further increase of pressure above ca. 4 GPa can either induce transformation into a static disordered phase \cite{capit17a} or to a presumably orthorhombic phase, as here reported, where the short-range order is still preserved (sharp Raman peaks) but the optical emission shows clear signs of carrier localization effects compatible with an incipient amorphization (see Fig. \ref{PL vs P}a). The first question is what triggers amorphization? Obviously, the MA cations cannot be, since they are locked and ordered throughout the crystal structure. By combining density functional theory and ab-initio molecular dynamics calculations, \cite{yixxx22a} an answer to this question has been recently provided for CsPbI$_3$, although valid in general for MHPs. Essentially, high pressure induces a phase instability driven by the softening of lattice vibrational modes associated with the tilting of the PbI$_6$ octahedrons, i.e. with a strong reduction of the Pb-I-Pb bond angle. The deformation caused by the stark octahedral tilting starts at different seeding points across the sample and extends gradually, leading to a lost of the long and short-range crystalline order. If so, the next question that arises is why is the onset of amorphization so sample dependent (reports from 2 to 7 GPa)? One possibility might be the different experimental conditions, for the phase behavior of MHPs under compression can be very sensitive to the degree of hydrostaticity of the pressure transmitting medium used. \cite{zhang17a} However, we propose an alternative explanation based on the amount of vacancies (mainly Pb vacancies) present in the sample. In a recent study on the optical emission of FA$_x$MA$_{1-x}$PbI$_3$ mixed crystals \cite{franc21a}, we have shown that the most common shallow defects in single crystalline MHPs are vacancies, mainly of Pb but also of the halogen and A-site cation. In view of the fact that the crystal structure is already deformed around a vacancy, we can foresee vacancies acting as seeding points to trigger the proposed pressure-induced lattice instability, leading to static disorder. \cite{yixxx22a}
\vspace{0.25cm}

We finally turn to a key result that concerns the temperature and pressure dependence of the N-H symmetric stretching vibration [$\nu_s$(NH$_3^+$)] of the MA cations, as determined by Raman scattering. This vibrational mode corresponds to the strongest peak in the Raman spectrum of MAPbBr$_3$ in the spectral range of the N-H stretching vibrations around 3000 cm$^{-1}$, as shown in the inset to Fig. \ref{N-H stretch}. This vibrational mode provides direct information about the coupling between the inorganic cage and the A-site cations, in particular, allowing one to unravel the conditional weight of H bonding and steric hindrance. This is so because the frequency of this vibrational mode shifts up or down upon changes in temperature, pressure or composition, if the coupling between both sublattices is dominated by steric or H-bonding effects, respectively. The frequency of the $\nu_s$(NH$_3^+$) vibration is determined by the strength of the covalent bond between nitrogen and hydrogen. In the H-bonding case, the electrostatic attraction between the H$^+$ and the negative halide ion of the inorganic cage weakens the N-H bond by elongating it, thus causing a redshift. \cite{wolff90a} On the contrary, the DSI is repulsive and stronger the closer the H and the halide atom become, which in turn shortens the N-H bond, causing a blueshift. As shown in Fig. \ref{N-H stretch}a, the frequency of the N-H stretching vibration decreases slightly with decreasing temperature (about 2 cm$^{-1}$ from room temperature down to 80 K). This is a clear indication that the H-bonding increases in importance with decreasing temperature, while the gradual cooling of the MA dynamics diminishes the steric effects. In fact, in the low-temperature orthorhombic phase, H-bonding is crucial to determine the arrangement (position and orientation) of the MA molecules within the cage voids. \cite{yixxx22a} However, at room temperature the application of moderate pressure causes a strong increase in frequency of the N-H stretching mode (ca. 8 cm$^{-1}$ up to 1.2 GPa), as displayed in Fig. \ref{N-H stretch}b. This is compelling evidence that, when the MA dynamics is fully unfolded, the DSI dominates the inter-sublattice coupling. We point out that the prominent role of DSI at ambient conditions is also demonstrated by the theoretically predicted and experimentally assessed blueshift of the $\nu_s$(NH$_3^+$) vibration for a reduction of the lattice parameter by the substitution of the halide atom from I to Br to Cl. \cite{leguy16a} This idea gathers additional support from a recent study that combines Raman scattering and density functional theory, where the entire vibrational spectrum of {\it isolated} MA$^+$ and FA$^+$ molecules is compared with that of MAPbX$_3$ and FAPbX$_3$ (X=I and Br), respectively. \cite{ibace22a} This comparison clearly shows that there are no hydrogen bonds in MHPs at room temperature.
\vspace{0.25cm}

\section{Discussion}

\begin{figure}[ht]
  \includegraphics[width=10cm,clip]{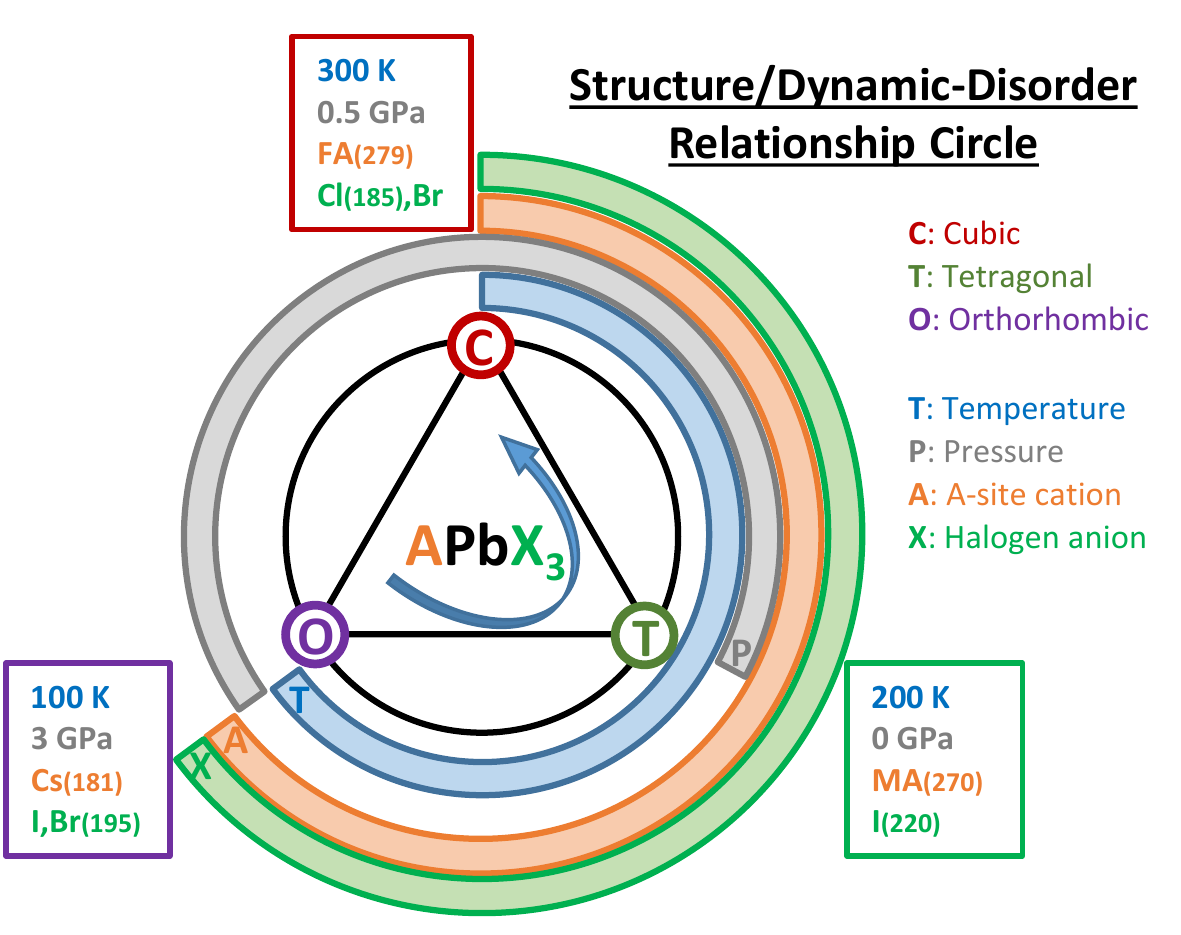}
  \caption{Schematic representation of the relationship between crystal structure and dynamic disorder in lead halide perovskites with formula APbX$_3$ as a function of temperature, pressure and composition (A-site cation and halogen anion types). The arrow pointing counterclockwise represents the direction of increase of the DSI. Numbers in parentheses correspond to the cation/anion ionic radii from the literature (in pm). \cite{travi16a,salib16a}
  }
  \label{S-D circle}
\end{figure}

At this point, it is worth offering a general discussion on the relationship between the crystal structures adopted by MHPs and the magnitude of dynamic disorder, as a function of different important parameters such as temperature, pressure and composition. For this purpose we use the sketch of Fig. \ref{S-D circle}, which serves both as a guide for discussion and as graphical summary of the "take-home" message of this work. The main hypothesis is that the strength of the dynamic steric interaction and, hence its leading role in the coupling between inorganic metal-halide network and the A-site cation sublattice, increases in direct proportion to the amount of dynamic disorder caused by the unfolded A-site cation dynamics. In this sense, the arrows in Fig. \ref{S-D circle} indicate the direction of increase of the DSI linked to the variation of the corresponding parameter.
\vspace{0.25cm}

We first discuss the effect of temperature, represented by the blue circle in Fig. \ref{S-D circle}. With increasing temperature, the structural sequence exhibited by MHPs is typically: orthorhombic (O)$\rightarrow$tetragonal (T)$\rightarrow$cubic (C). Concomitantly with the thermal activation of vibrations, rotations and translations of the A-site cations within the cage voids, there is an increase in dynamic disorder and, thus, of the DSI. The increase in entropy from dynamic disorder overcompensates both the decrease in structural entropy for the more symmetric structures and the detrimental effect of the lattice thermal expansion  on the DSI.
\vspace{0.25cm}

We next consider the impact of replacing the A-site cation on the structural behavior of MHPs (orange circle in Fig. \ref{S-D circle}). The same O$\rightarrow$T$\rightarrow$C sequence is observed for an increasing cation size, when going from an atom like Cs to a molecule like MA and a larger one such as FA. Even though the lattice parameter of the perovskite increases slightly for larger A-site cation size, the effective volume filled by the A-site cation ($V_A$), the so-called steric bulk, increases faster than the void volume ($V_v$) itself. The volume $V_A$ can be inferred, for example, from the atomic displacement parameter plots at 50\% probability from neutron scattering \cite{welle15a,weber18a} or molecular dynamics calculations. \cite{ghosh17a,ghosh19a} The key point is that the strength of the DSI is proportional to the ratio $\frac{V_A}{V_v}$ and being repulsive in nature, the fast movement of the A-site cations produces the same effect of an {\it internal} pressure acting outwards on the imaginary walls of the cage voids. As clearly shown by molecular dynamics simulations, \cite{ghosh17a,ghosh19a} the spherical atomic-density cloud generated by the movement of the A-site cations in the three spatial directions favors a cubic void environment, thus stabilizing the cubic phase. In contrast, a free movement solely in a plane favors the stabilization of the tetragonal phase, whereas the orthorhombic phase is only compatible with the locking of the A-site cations inside the cage voids. A nice example of the effect of the A-site cation size can be appreciated for the series of lead bromide compounds. \cite{manni20a,naqvi22a}
\vspace{0.25cm}

We now turn to the discussion of the effects of the halogen atom substitution, which are illustrated by the green circle in Fig. \ref{S-D circle}. In this case, the effects on the structural behavior are more subtle than for the preceding parameters. However, one can recognize certain correlation between the structural sequence O$\rightarrow$T$\rightarrow$C and the reduction of the ionic radius of the halogen atoms. The heavier the halogen atom, the larger its ionic radius, which leads to an increase of the lattice parameter, i.e. of $V_v$. This means that the DSI decreases with increasing ionic radius of the halogens, which explains why chlorine compounds are more prone to stabilize in the cubic phase than the bromide and iodide counterparts. At least, the T$\rightarrow$C transition temperature, for example for the MAPbX$_3$ family, shifts to higher temperatures as the halogen ionic radius increases. \cite{leguy16a} As shown below for the case of varying the pressure, halogen substitution works in a similar way as what is known as {\it chemical pressure}.
\vspace{0.25cm}

Finally, we discuss the structural effect of an external hydrostatic pressure. The corresponding grey circle in Fig. \ref{S-D circle} appears out of phase with respect to the others, because the observed structural sequence under compression is T$\rightarrow$C$\rightarrow$O, as for the emblematic case of MAPbI$_3$. \cite{szafr16a,franc18a,capit16a,jaffe16a} This seems a priori counterintuitive. In fact, since the effect on the void volume $V_v$ of thermal expansion is opposite to that of compression, one would expect the pressure effect to be represented in the sketch of Fig. \ref{S-D circle} by a similar circle as for the temperature but going clockwise instead of anticlockwise. The only way to understand such behavior is by considering the DSI as the dominant interaction against H bonding. As mentioned before, when the A-site cation dynamics is fully unfolded and due to the repulsive character of the DSI, the moving A-site cations exert an outward force to the surrounding octahedrons, which partly counteracts the effect of the applied pressure. In the tetragonal phase, stable at ambient conditions for MAPbI$_3$, where the dynamics of the MA cations is restricted to the tetragonal plane, such a reaction of the MA molecules to compression is only expected along the (a,b) tetragonal axes. To the contraction of the tetragonal axes under pressure follows a reaction of the moving MA cations, mediated by DSI, that repels the tilted octahedrons slowing down further pressure-induced tilting. This leads to an effective asymmetry in the compressibility of the inorganic cage, in view of the fact that the longer (unperturbed) c axis would be more compressible than the (distorted) tetragonal ones. Thus, with increasing pressure the tetragonal distortion diminishes up to the point where the compressed crystal structure is almost cubic. At that moment is when the MA dynamics becomes unleashed in all three directions in space, what in turn stabilizes the cubic phase at finite pressure. \cite{ghosh17a,ghosh19a} Further compression will eventually induce a transformation into an orthorhombic phase, which is thermodynamically more stable at reduced volumes and after the MA dynamics has collapsed. This phenomenology is a unique signature of the DSI present in MHPs. We point out that from a pure structural point of view, ferroelectric perovskites like CsGeX$_3$ with X=I, Br, and Cl also exhibit a similar behavior under pressure. \cite{chenx23a,schwa96a} However, the reason for it is the pressure-induced reduction up to a full collapse of the Jahn-Teller distortion giving rise to the ferroelectric polarization. Lead halide perovskites, in contrast, are not ferroelectric but ferroelastic \cite{ambro22a} and the transformation from tetragonal to cubic structure under compression is the consequence of a gradual pressure-induced, DSI aided reduction of the tetragonal symmetry.
\vspace{0.25cm}

\section{Conclusion}
In summary, we have performed a systematic study of the optical emission and vibrational properties of single crystalline MAPbBr$_3$ as a function of temperature and hydrostatic pressure using photoluminescence and Raman scattering spectroscopy. These results combined with the available literature data on other closely-related MHPs allowed us to unravel the underlying physics relating the crystal structure stability, depending on composition as well as temperature and pressure conditions, and the dynamic disorder caused by the fast A-site cation dynamics. The main finding is that a full understanding of the relationship between structure and dynamic disorder in MHPs can only be achieved if dynamic steric effects are taken into account; H-bonding alone is insufficient. The leitmotif for the observed trends regarding the crystal phase sequences obtained with increasing temperature, pressure and A-site cation size or with decreasing halogen ionic radius is a strengthening of the DSI, which is directly linked with the magnitude of the dynamic disorder induced by the unfolded A-site cation roto-translational dynamics. Furthermore, we offer an explanation for the large spread in the reported values of the onset of the pressure-induced amorphization or static-disordered state, ubiquitous in MHPs. Here we suggest that vacancies (mainly of lead) act as seeding points for the pressure-induced lattice instability due to the softening of phonon modes related to octahedral tilting, as proposed to trigger amorphization. \cite{yixxx22a} Since the lattice is already deformed at a vacancy, the number of vacancies would then determine the onset of amorphization, making its observation fairly sample dependent. In this way, we believe to have deepened our understanding of a very fundamental issue for MHPs, namely the crystal-structure/dynamic-disorder relationship, thus contributing to advance the development of optoelectronic applications of this exceptional class of materials.

% Experimental section

\section{Methods}

%\threesubsection{Growth of the MAPbBr$_3$ single crystals}
\subsection{Growth of the MAPbBr$_3$ single crystals}

The inverse solubility method of Saidaminov \textit{et al.} \cite{saida15a} was developed to produce crystals of MAPbBr$_{3}$. Stoichiometric quantities of MABr (GreatCell Solar) and PbBr$_{2}$ (Merck, 99\%) were dissolved at 20 $^\circ$C in dry dimethylformamide (Alfa Aesar). When fully dissolved, the solution was heated to 80 $^\circ$C and left undisturbed for 3 hours to allow crystallisation. The remaining solution was filtered off and large single crystals were oven dried at 100 $^\circ$C overnight.

\subsection{High-pressure experiments}

The high-pressure photoluminescence and micro-Raman scattering measurements were performed at room temperature employing a gasketed diamond anvil cell (DAC). Anhydrous propanol was used as pressure transmitting medium which ensures good hydrostatic conditions in the pressure range of the present experiments (perfectly hydrostatic up to 4.2 GPa \cite{angel07a}) and proved chemically inert to MAPbBr$_3$.
For loading the DAC, small chips with a thickness below ca. 30 $\mu$m were produced by crushing a big MAPbBr$_3$ single crystal between two glass slides. By close inspection of the debris we were able to pick up small enough, good-quality single crystals, recognized by their flat and shiny surface under the microscope. This simple but effective procedure allowed us to avoid the thinning of the sample by either mechanical polishing or chemical etching, which are known to spoil the quality of such soft crystals. Small pieces of about 100$\times$100 $\mu$m$^2$ in size were placed into the DAC together with a ruby sphere for pressure calibration \cite{pressure2}. Here we point out that the Inconel gasket was intentionally pre-indented to a fairly large thickness of 120 $\mu$m, before drilling a hole of ca. 250 $\mu$m with a spark-gap machine from EasyLab. The reason was to be able to adjust the pressure with the DAC in steps less than 0.05 GPa, mainly at very low pressures (below 1 GPa). For this purpose an electric motor drive was used to change the pressure in a continuous manner and at low speed (by ca. 0.05 GPa/min). In return, the maximum pressure reached in our experiments was about 7 GPa. Regarding the high accuracy claimed in the measurement of the pressure, we point out that we always loaded more than one ruby sphere into the DAC for a multi-point determination of the pressure. The excitation of the ruby fluorescence was performed using extremely low laser powers in the range of a few tens of nW, in order to avoid any heating-induced shift of the ruby emission. Furthermore, the pressure was determined immediately before and after each PL or Raman measurement, to account for effects of mechanical relaxation of the DAC upon changes in pressure. The temperature of the room was also frequently monitored to correct for an eventual temperature increase of the room, for example, from the morning to the evening or if another heat-generating equipment (laser, vacuum pump, etc.) was switched on nearby. Finally, the backlash of the spectrometer was also considered and to minimize its effect, the fluorescence of the ruby was measured, acquiring the spectrum by forcing the spectrometer to approach its final position always from the same side.
\vspace{0.25cm}

\subsection{PL \& Raman measurements}

For the high-pressure experiments, the PL spectra were excited with the 405 nm line of laser diode, whereas for the PL measurements at low temperatures the 514.5 nm line of an Ar$^+$-ion laser was employed, using a very low incident light power below 2 $\mu$W. The latter was selected as the closest available laser line to the MAPbBr$_3$ gap. This turned out to be very important to attain long time stability and reproducibility of the PL emission by reducing as much as possible laser heating effects due to thermalization of photo-generated hot carriers. For the Raman measurements either an infrared diode laser emitting at 785 nm or the 633 nm line of a He-Ne laser was employed for excitation of the low-frequency spectra (below 500 cm$^{-1}$) and the high-frequency ones (around 3.000 cm$^{-1}$), respectively. The former turned out most suitable to excite the vibrational modes of the inorganic cage, providing also the highest spectral resolution and stray-light rejection. In all cases, a very low incident light power density below 15 W/cm$^2$ was used to avoid any photo-degradation of the samples, such that thermal damage by the laser can be safely ruled out. Spectra were collected using a 20$\times$ long working distance objective with NA=0.35 and dispersed with a high-resolution LabRam HR800 grating spectrometer equipped with a charge-coupled device detector. PL spectra were corrected for the spectral response of the spectrometer by normalizing each spectrum using the detector and the 600-grooves/mm grating characteristics. Temperature-dependent measurements on large single crystals exhibiting flat surfaces were carried out between 80 and 320 K using a gas flow cryostat from CryoVac with optical access that fits under the microscope of the LabRam setup.
\vspace{0.25cm}

\medskip
\textbf{Supporting Information} \par %Please delete the Suppporting Information statement if it is not applicable. Please supply Supporting Information in another file. Supporting information should not be provided in .tex format
The Supporting Information contains a set of PL spectra recorded at five different temperatures in the range of approx. 10 to 50 K for each of the ten compositions of the FA$_x$MA$_{1-x}$PbI$_3$ system studied here, showing details of the lineshape fits performed to the PL spectra using multiple Gaussian-Lorentzian cross-product functions. The results of the line-shape fits concerning the peak energy, line width and intensity for the main emission features, plotted as a function of temperature, are also included.

% Acknowledgements
\medskip
\textbf{Acknowledgements} \par %delete if not applicable))
The Spanish "Ministerio de Ciencia e Innovaci\'{o}n (MICINN)" is gratefully acknowledged for its support through grant CEX2019-000917-S (FUNFUTURE) in the framework of the Spanish Severo Ochoa Centre of Excellence program and the AEI/FEDER(UE) grants PGC2018-095411-B-100 (RAINBOW) and PID2021-128924OB-I00 (ISOSCELLES). The authors also thank the Catalan agency AGAUR for grant 2017-SGR-00488 and the National Network "Red Perovskitas" (MICINN funded). K.X. acknowledges a fellowship (CSC201806950006) from China Scholarship Council and the PhD programme in Materials Science from Universitat Aut\`{o}noma de Barcelona in which he was enrolled. B.C. thanks the EPSRC for PhD studentship funding via the University of Bath, CSCT CDT (EP/G03768X/1).

\medskip
\textbf{Author contributions}
Conceptualization: A.R.G.; experiments-data generation: K.X. and L.P.-F.; materials synthesis: B.C.; analysis: K.X., L.P.-F. and A.R.G.; supervision: M.I.A., M.T.W. and A.R.G.; writing-original draft preparation: A.R.G. All authors reviewed the manuscript.

\medskip
\textbf{Data availability}
All data generated or analysed during this study are either included in this published article and its supplementary information files or are available from the corresponding author on reasonable request.

% References
\medskip

% Use the following code if you wish to generate your bibliography with BibTeX;
% replace the string "MSP-template" below with the name(s) of
% the BibTeX data base(s) you want to use.
% The resulting bibliography-output (the content of the .bbl file)
% must be pasted back into this file before submission.
% Please also include your BibTeX data base file(s) in your submission
% so that we can re-run BibTeX if necessary.
%
%\bibliographystyle{MSP}
%\bibliography{MSP-template}

\begin{thebibliography}{99}

\bibitem{NREL} See {\it Best Research-Cell Efficiency Chart}, source: National Renewable Energy Laboratory (NREL), Golden, Colorado, USA. www.nrel.gov/pv/cell-efficiency.html (2022).
\bibitem{golds26a} Goldschmidt, V. M. Die Gesetze der Krystallochemie, {\it Die Naturwissenschaften} {\bf 14}, 477-485 (1926).
\bibitem{frost16a} Frost, J. M., and Walsh, A. What is moving in hybrid halide perovskite solar cells? {\it Acc. Chem. Res.} {\bf 49}, 528-535 (2016).
\bibitem{bakul15a} Bakulin, A. A., Selig, O., Bakker, H. J., Rezus, Y. L., M\"{u}ller, C. Glaser, T., Lovrincic, R., Sun, Z., Chen, Z., Walsh, A., et al. Real-time observation of organic cation reorientation in methylammonium lead iodide perovskites. {\it J. Phys. Chem. Lett.} {\bf 6}, 3663-3669 (2015).
\bibitem{selig17a} Selig, O., Sadhanala, A., M\"{u}ller, C., Lovrincic, R., Chen, Z., Rezus, Y. L., Frost, J. M., Jansen, T. L., Bakulin, A. A. Organic cation rotation and immobilization in pure and mixed methylammonium lead-halide perovskites. {\it J. Am. Chem. Soc.} {\bf 139}, 4068-4074 (2017).
\bibitem{welle15a} Weller, M. T., Weber, O. J., Henry, P. F., Di Pumpo, A. M., Hansen, T. C. Complete structure and cation orientation in the perovskite photovoltaic methylammonium lead iodide between 100 and 352 K. {\it Chem. Commun.} {\bf 51}, 4180-4183 (2015).
\bibitem{weber18a} Weber, O. J., Ghosh, D., Gaines, S., Henry, P. F., Walker, A. B., Islam, M. S., Weller, M. T. Phase behavior and polymorphism of formamidinium lead iodide. {\it Chem. Mater.} {\bf 30}, 3768-3778 (2018).
\bibitem{szafr16a} Szafra\'{n}ski, M., Katrusiak, A. Mechanism of pressure-induced phase transitions, amorphization, and absorption-edge shift in photovoltaic methylammonium lead iodide. {\it J. Phys. Chem. Lett.} {\bf 7}, 3458-3466 (2016).
\bibitem{ghosh17a} Ghosh, D., Atkins, P. W., Islam, M. S., Walker, A. B., Eames, C. Good vibrations: Locking of octahedral tilting in mixed-cation iodide perovskites for solar cells. {\it ACS Energy Lett.} {\bf 2}, 2424-2429 (2017).
\bibitem{ghosh19a} Ghosh, D., Aziz, A., Dawson, J. A., Walker, A. B., Islam, M. S. Putting the squeeze on lead iodide perovskites: Pressure-induced effects to tune their structural and optoelectronic behavior. {\it Chem. Mater.} {\bf 31}, 4063-4071 (2019).
\bibitem{maity23a} Maity, S., Verma, S., Ramaniah, L. M., Srinivasan, V. Deciphering the nature of temperature-induced phases of MAPbBr$_3$ by ab initio molecular dynamics. {\it Chem. Mater.} {\bf 34}, 10459-10469 (2022). %DOI: 10.1021/acs.chemmater.2c02453
\bibitem{matto15a} Mattoni, A., Filippetti, A., Saba, M., Delugas, P. Methylammonium rotational dynamics in lead halide perovskite by classical molecular dynamics: the role of temperature. {\it J. Phys. Chem. C} {\bf 119}, 17421-17428 (2015).
%\bibitem{H-bond}
%\bibitem{steric}
\bibitem{yinxx17a} Yin, T., Fang, Y., Fan, X., Zhang, B., Kuo, J.-L., White, T. J., Chow, G. M., Yan, J., and Shen, Z. X. Hydrogen-bonding evolution during the polymorphic transformations in CH$_3$NH$_3$PbBr$_3$: Experiment and theory. {\it Chem. Mater.} {\bf 29}, 5974-5981 (2017).
\bibitem{capit17a} Capitani, F., Marini, C., Caramazza, S., Dore, P., Pisanu, A., Malavasi, L., Nataf, L., Baudelet, F., Brubach, J.-B., Roy, P., Postorino, P. Locking of methylammonium by pressure-enhanced H-bonding in (CH$_3$NH$_3$)PbBr$_3$ hybrid perovskite. {\it J. Phys. Chem. C} {\bf 121}, 28125-28131 (2017).
\bibitem{yesud20a} Yesudhas, S., Burns, R., Lavina, B., Tkachev, S. N., Sun, J., Ullrich, C. A., Guha, S. Coupling of organic cation and inorganic lattice in methylammonium lead halide perovskites: Insights into a pressure-induced isostructural phase transition. {\it Phys. Rev. Mater.} {\bf 4}, 105403 (2020).
\bibitem{leexx16a} Lee, J.-H., Bristowe, N. C., Lee, J. H., Lee, S.-H., Bristowe, P. D., Cheetham, A. K., Jang, H. M. Resolving the physical origin of octahedral tilting in halide perovskites. {\it Chem. Mater.} {\bf 28}, 4259-4266 (2016).
\bibitem{leexx16b} Lee, J. H., Lee, J.-H., Kong, E.-H., Jang, H. M. The nature of hydrogen-bonding interaction in the prototypic hybrid halide perovskite, tetragonal CH$_3$NH$_3$PbI$_3$. {\it Sci. Rep.} {\bf 6}, 21687 (2016).
\bibitem{brivi15a} Brivio, F., Frost, J. M., Skelton, J. M., Jackson, A. J., Weber, O. J., Weller, M. T., Go\~{n}i, A. R., Leguy, A. M. A., Barnes, P. R. F., Walsh, A. Lattice dynamics and vibrational spectra of the orthorhombic, tetragonal, and cubic phases of methylammonium lead iodide. {\it Phys. Rev. B} {\bf 92}, 144308/1-8 (2015).
\bibitem{leguy16a} Leguy, A. M. A., Go\~{n}i, A. R., Frost, J. M., Skelton, J., Brivio, F., Rodr\'{\i}guez-Mart\'{\i}nez, X., Weber, O. J., Pallipurath, A., Alonso, M. I., Campoy-Quiles, M., Weller, M. T., Nelson, J., Walsh, A., Barnes, P. R. F. Dynamic disorder, phonon lifetimes, and the assignment of modes to the vibrational spectra of methylammonium lead halide perovskites. {\it Phys. Chem. Chem. Phys.} {\bf 18}, 27051-27066 (2016).
\bibitem{sharm20a} Sharma, R., Menahem, M., Dai, Z., Gao, L., Brenner, T. M., Yadgarov, L., Zhang, J., Rakita, Y., Korobko, R., Pinkas, I., Rappe, A. M., Yaffe, O. Lattice mode symmetry analysis of the orthorhombic phase of methylammonium lead iodide using polarized Raman. {\it Phys. Rev. Mater.} {\bf 4}, 051601R (2020).
\bibitem{franc18a} Francisco-L\'{o}pez, A., Charles, B., Weber, O. J., Alonso, M. I., Garriga, M., Campoy-Quiles, M., Weller, M. T., Go\~{n}i, A. R. Pressure-induced locking of methylammonium cations versus amorphization in hybrid lead iodide perovskites. {\it J. Phys. Chem. C} {\bf 122}, 22073-22082 (2018).
\bibitem{franc20a} Francisco-L\'{o}pez, A., Charles, B., Alonso, M. I., Garriga, M., Campoy-Quiles, M., Weller, M. T., Go\~{n}i, A. R. Phase diagram of methylammonium/formamidinium lead iodide perovskite solid solutions from temperature-dependent photoluminescence and Raman spectroscopies. {\it J. Phys. Chem. C} {\bf 124}, 3448-3458 (2020).
\bibitem{pogli87a} Poglitsch, A., and Weber, D. Dynamic disorder in methylammoniumtrihalogenoplumbates (II) observed by millimeter-wave spectroscopy. {\it J. Chem. Phys.} {\bf 87}, 6373-6378 (1987).
\bibitem{onoda90a} Onoda-Yamamuro, N., Matsuo, T., and Suga, H. Calorimetric and IR spectroscopic studies of phase transitions in methylammonium trihalogenoplumbates (II). {\it J. Phys. Chem. Solids} {\bf 51}, 1383-1395 (1990).
\bibitem{onoda92a} Onoda-Yamamuro, N., Matsuo, T., and Suga, H. Dielectric study of CH$_3$NH$_3$PbX$_3$ (X = Cl, Br, I). {\it J. Phys. Chem. Solids} {\bf 53}, 935-939 (1992).
\bibitem{capit16a} Capitani, F., Marini, C., Caramazza, S., Postorino, P., Garbarino, G., Hanfland, M., Pisanu, A., Quadrelli, P., Malavasi, L. High-pressure behavior of metylammonium lead iodide (MAPbI$_3$) hybrid perovskite. {\it J. Appl. Phys.} {\bf 119}, 185901/1-6 (2016).
\bibitem{wangx17a} Wang, T., Daiber, B., Frost, J. M., Mann, S. A., Garnett, E. C., Walsh, A., and Ehrler, B. Indirect to direct bandgap transition in methylammonium lead halide perovskite. {\it Energy Environ. Sci.} {\bf 10}, 509-515 (2017).
\bibitem{jaffe16a} Jaffe, A., Lin, Y., Beavers, C. M., Voss, J., Mao, W. L., Karunadasa, H. I. High-pressure single-crystal structures of 3D lead-halide hybrid perovskites and pressure effects on their electronic and optical properties. {\it ACS Cent. Sci.} {\bf 2}, 201-209 (2016).
\bibitem{ouxxx16a} Ou, T., Yan, J., Xiao, C., Shen, W., Liu, C., Liu, X., Han, Y., Ma, Y., Gao, C. Visible light response, electrical transport, and amorphization in compressed organo lead iodine perovskites. {\it Nanoscale} {\bf 8}, 11426-11431 (2016).
\bibitem{jiang16a} Jiang, S., Fang, Y., Li, R., Xiao, H., Crowley, J., Wang, C., White, T. J., Goddard III, W. A., Wang, Z., Baikie, T., Fang, J. Pressure-dependent polymorphism and band-gap tuning of methylammonium lead iodide perovskite. {\it Angew. Chem. Int. Ed.} {\bf 55}, 6540-6544 (2016).
\bibitem{kongx16a} Kong, L., Liu, G., Gong, J., Hu, Q., Schaller, R. D., Dera, P., Zhang, D., Liu, Z., Yang, W., Tang, Y., Wang, C., Wei, S.-H., Xu, T., Mao, H.-K. Simultaneous band-gap narrowing and carrier-lifetime prolongation of organic-inorganic trihalide perovskites. {\it PNAS} {\bf 113}, 8910-8915 (2016).
\bibitem{swain07a} Swainson, I. P., Tucker, M. G., Wilson, D. J., Winkler, B., Milman, V. Pressure response of an organic-inorganic perovskite: Methylammonium lead bromide. {\it Chem. Mater.} {\bf 19}, 2401-2405 (2007).
\bibitem{manni20a} Mannino, G., Deretzis, I., Smecca, E., La Magna, A., Alberti, A., Ceratti, D., Cahen, D. Temperature-dependent optical band gap in CsPbBr$_3$, MAPbBr$_3$, and FAPbBr$_3$ single crystals. {\it J. Phys. Chem. Lett.} {\bf 11}, 2490-2496 (2020).
\bibitem{wangx15a} Wang, Y., L\"{u}, X., Yang, W., Wen, T., Yang, L., Ren, X., Wang, L., Lin, Z., and Zhao, Y. Pressure-induced phase transformation, reversible amorphization, and anomalous visible light response in organolead bromide perovskite. {\it J. Am. Chem. Soc.} {\bf 137}, 11144-11149 (2015).
\bibitem{wangx16b} Wang, L., Wang, K., Xiao, G., Zeng, Q., Zou, B. Pressure-induced structural evolution and band gap shifts of organometal halide perovskite-based methylammonium lead chloride. {\it J. Phys. Chem. Lett.} {\bf 7}, 5273-5279 (2016).
\bibitem{carpe23a} Carpenella, V., Ripanti, F., Stellino, E., Fasolato, C., Nucara, A., Petrillo, C., Malavasi, L., Postorino, P. High pressure behavior of $\delta$-phase of formamidinium lead iodide by optical spectroscopy. {\it J. Phys. Chem. C} (2023). DOI: 10.1021/acs.jpcc.2c08253
\bibitem{wangx16a} Wang, L., Wang, K., Zou, B. Pressure-induced structural and optical properties of organometal halide perovskite-based formamidinium lead bromide. {\it J. Phys. Chem. Lett.} {\bf 7}, 2556-2562 (2016).
\bibitem{mohan19a} Mohanty, A., Swain, D., Govinda, S., Row, T. N. G., Sarma, D. D. Phase diagram and dielectric properties of MA$_{1-x}$FA$_x$PbI$_3$. {\it ACS Energy Lett.} {\bf 4}, 2045-2051 (2019).
\bibitem{sharm21a} Sharma, V. K., Mukhopadhyay, R., Mohanty, A., Garc\'{\i}a Sakai, V., Tyagi, M., and Sarma, D. D. Contrasting effects of FA substitution on MA/FA rotational dynamics in FA$_x$MA$_{1-x}$PbI$_3$. {\it J. Phys. Chem. C} {\bf 125}, 13666-13676 (2021).
\bibitem{yixxx22a} Yi, S., and Lee, J.-H. Degenerate lattice-instability-driven amorphization under compression in metal halide perovskite CsPbI$_3$. {\it J. Phys. Che. Lett.} {\bf 13}, 9449-9455 (2022).
\bibitem{lixxx20a} Li, M., Liu, T., Wang, Y., Yang, W., and L\"{u}, X. Pressure responses of halide perovskites with various compositions, dimensionalities, and morphologies. {\it Matter Radiat. Extremes} {\bf 5}, 018201 (2020).
\bibitem{celes22a} Celeste, A., and Capitani, F. Hybrid perovskites under pressure: Present and future directions. {\it J. Appl. Phys.} {\bf 132}, 220903 (2022).
\bibitem{galco17a} Galkowski, K., Mitioglu, A. A., Surrente, A., Yang, Z., Maude, D. K., Kossaki, P., Eperon, G. E., Wang, J. T.-W., Snaith, H. J., Plochocka, P., Nicholas, R. J. Spatially resolved studies of the phases and morphology of methylammonium and formamidinium lead tri-halide perovskites. {\it Nanoscale} {\bf 9}, 3222-3230 (2017).
\bibitem{tilch16a} Tilchin, J., Dirin, D. N., Maikov, G. I., Sashchiuk, A., Kovalenko, M. V., Lifshitz, E. Hydrogen-like Wannier-Mott excitons in single crystal of methylammonium lead bromide perovskite. {\it ACS Nano} {\bf 10}, 6363-6371 (2016).
\bibitem{franc21a} Francisco-L\'{o}pez, A., Charles, B., Alonso, M. I., Garriga, M., Weller, M. T., Go\~{n}i, A. R. Photoluminescence of bound-exciton complexes and assignment to shallow defects in methylammonium/formamidinium lead iodide mixed crystals. {it Adv. Optical Mater.}, 2001969/1-9 (2021).
\bibitem{franc19a} Francisco L\'{o}pez, A., Charles, B., Weber, O. J., Alonso, M. I., Garriga, M., Campoy-Quiles, M., Weller, M. T., Go\~{n}i, A. R. Equal footing of thermal expansion and electron-phonon interaction in the temperature dependence of lead halide perovskite band gaps. {\it J. Phys. Chem. Lett.} {\bf 10}, 2971-2977 (2019).
\bibitem{gonix98a} Go\~{n}i, A. R., and Syassen, K. Optical properties of semiconductors under pressure. {\it Semicond. Semimetals} {\bf 54}, 247-425, (1998) and references therein.
\bibitem{frost14a} Frost, J. M.  Butler, K. T., Brivio, F., Hendon, C. H., van Schilgaarde, M., Walsh, A. Atomistic origins of high-performance in hybrid halide perovskite solar cells. {\it Nano Lett.} {\bf 14}, 2584-2590 (2014).
\bibitem{evenx15a} Even, J., Pedesseau, L., Katan, C., Kepenekian, M., Lauret, J.-S., Sapori, D., Deleporte, E. Solid-state physics perspective on hybrid perovskite semiconductors. {\it J. Phys. Chem. C} {\bf 119}, 10161-10177 (2015).
\bibitem{yaffe17a} Yaffe, O., Guo, Y., Tan, L. Z., Egger, D. A., Hull, T., Stoumpos, C. C., Zheng, F., Heinz, T. F., Kronik, L., Kanatzidis, M. G., Owen, J. S., Rappe, A. M., Pimenta, M. A., Brus, L. E. Local polar fluctuations in lead halide perovskite crystals. {\it Phys. Rev. Lett.} {\bf 118}, 136001/1-6 (2017).
\bibitem{sendn16a} Sendner, M., Nayak, P. K., Egger, D. A., Beck, S., M\"{u}ller, C., Epding, B., Kowalsky, W., Kronik, L., Snaith, H. J., Puccia, A., Lovrincic, R. Optical phonons in methylammonium lead halide perovskites and implications for charge transport. {\it Mater. Horiz.} {\bf 3}, 613-620 (2016).
\bibitem{tusch18a} Tuschel, D. Exploring resonance Raman spectroscopy. {\it Spectros.} {\bf 33}, 12-19 (2018).
\bibitem{reuve22a} Reuveni, G., Diskin-Posner, Y., Gehrmann, C., Godse, S., Gkikas, G. G., Buchine, I., Aharon, S., Korobko, R., Stoumpos, C. C., Egger, D. A., and Yaffe, O. Static and dynamic disorder in formamidinium lead bromide single crystals. arXiv:2211.06904v1 [cond-mat.mtrl-sci] 13 Nov 2022.
\bibitem{zhang17a} Zhang, R., Cai, W., Bi, T., Zarifi, N., Terpstra, T., Zhang, C., Verdeny, Z. V., Zurek, E., and Deemyad, S. Effects of nonhydrostatic stress on structural and optoelectronic properties of methylammonium lead bromide perovskite. {\it J. Phys. Chem. Lett.} {\bf 8}, 3457-3465 (2017).
\bibitem{wolff90a} Wolff, H., Schmidt, U., and Wolff, E. NH$_2$ stretching vibration absorption and association mechanism of methylamine in $n$-hexane and carbon tetrachloride. {\it Spectrochimica Acta} {\bf 46A}, 85-89 (1990).
\bibitem{ibace22a} Ibaceta-Ja\~{n}a, J., Chugh, M., Novikov, A. S., Mirhosseini, H., K\"{u}hne, T. D., Szyszka, B., Wagner, M. R., and Muydinov, R. Do lead halide hybrid perovskites have hydrogen bonds? {\it J. Phys. Chem. C} {\bf 126}, 16215-16226 (2022).
\bibitem{travi16a} Travis, W., Glover, E. N. K., Bronstein, H., Scanlon, D. O., and Palgrave, R. G. On the application of the tolerance factor to inorganic and hybrid halide perovskites: a revised system. {\it Chem. Sci.} {\bf 7}, 4548-4556 (2016).
\bibitem{salib16a} Saliba, M., Matsui, T., Seo, J.-Y., Domanski, K., Correa-Baena, J.-P., Nazeeruddin, M. K., Zakeeruddin, S. M., Tress, W., Abate, A., Hagfeldt, A., and Gr\"{a}tzel, M. Cesium-containing triple cation perovskite solar cells: improved stability, reproducibility and high efficiency. {\it Energy Environ. Sci.} {\bf 9}, 1989-1997 (2016).
\bibitem{naqvi22a} Naqvi, F. H., Ko, J.-H., Kim, T. H., Ahn, C. W., Hwang, Y., Sheraz, M., Kim, S. A-site cation effect on optical phonon modes and thermal stability in lead-based perovskite bromide single crystals using Raman spectroscopy. {\it J. Korean Phys. Soc.} {\bf 81}, 230-240 (2022).
\bibitem{chenx23a} Chen, R., Liu, C., Chen, Y., Ye, C., Chen, S., Cheng, J., Cao, S., Wang, S., Cui, A., Hu, Z., Lin, H., Wu, J., Kong, X. Y., and Ren, W. Ferroelectric CsGeI$_3$ single crystals with a perovskite structure grown from aqueous solution. {\it J. Phys. Chem. C} (2023). DOI:10.1021/acs.jpcc.2c06818
\bibitem{schwa96a} Schwarz, U., Wagner, F., Syassen, K., and Hillebrecht, H. Effect of pressure on the optical-absorption edges of CsGeBr$_3$ and CsGeCl$_3$. {\it Phys. Rev. B} {\bf 53}, 12545-12548 (1996).
\bibitem{ambro22a} Ambrosio, F., De Angelis, F., and Go\~{n}i, A. R. The ferroelectric-ferroelastic debate about metal halide perovskites. {\it J. Phys. Chem. Lett.} {\bf 13}, 7731-7740 (2022).
\bibitem{saida15a} M. I. Saidaminov, A. L. Abdelhady, B. Murali, E. Alarousu, V. M. Burlakov, W. Peng, I. Dursun, L. Wang, Y. He, G. Maculan, A. Goriely, T. Wu, O. F. Mohammed, O. M. Bakr, High-quality bulk hybrid perovskite single crystals within minutes by inverse temperature crystallization. {\it Nat. Commun.} {\bf 6}, 7586 (2015).
\bibitem{angel07a} Angel, R. J., Bujak, M., Zhao, J., Gatta, G. D., Jacobsen, S. D. Effective hydrostatic limits of pressure media for high-pressure crystallographic studies, {\it J. Appl. Crystallogr.} {\bf 40}, 26-32 (2007).
\bibitem{pressure2} Mao, H.-K., Xu, J., Bell, P. M. Calibration of the ruby pressure gauge to 800 kbar under quasi-hydrostatic conditions. {\it J. Geophys. Res.} {\bf 91}, 4673-4676 (1986).


\end{thebibliography}

\textbf{References}\\

\end{document}